# Comparative Analysis of Human Mobility Patterns: Utilizing Taxi and Mobile (SafeGraph) Data to Investigate Neighborhood-Scale Mobility in New York City

Authors: Yuqin Jiang[1,2], Zhenlong Li[3], Joon-Seok Kim[4], Huan Ning[3], Su Yeon Han[1*]
[1] Department of Geography and Environmental Studies, Texas State University
[2] Department of Geography and Environment, University of Hawaiʻi at Mānoa
[3] Geoinformation and Big Data Research Laboratory, Department of Geography, Pennsylvania State University
[4] Computer Science, Emory University

[*] su.han@txstate.edu

## Abstract

Numerous researchers have utilized GPS-enabled vehicle data and SafeGraph mobility data to analyze human movements. However, the comparison of their ability to capture human mobility remains unexplored. This study investigates differences in human mobility using taxi trip records and the SafeGraph dataset in New York City neighborhoods. The analysis includes neighborhood clustering to identify population characteristics and a comparative analysis of mobility patterns. Our findings show that taxi data tends to capture human mobility to and from locations such as Lower Manhattan, where taxi demand is consistently high, while often underestimating the volume of trips originating from areas with lower taxi demand, particularly in the suburbs of NYC. In contrast, SafeGraph data excels in capturing trips to and from areas where commuting by driving one's own car is common, but underestimates trips in pedestrian-heavy areas. The comparative analysis also sheds new light on transportation mode choices for trips across various neighborhoods. The results of this study underscore the importance of understanding the representativeness of human mobility big data and highlight the necessity for careful consideration when selecting the most suitable dataset for human mobility research.

## 1. Introduction

The importance of mobile location datasets has surged, playing an increasingly important role in uncovering human mobility patterns in scientific research (Z. Li et al., 2024; Shaw et al., 2016). Nowadays, smartphones, equipped with high-accuracy Global Positioning System (GPS) receivers, seamlessly communicate with various applications, providing real-time locations (Hu et al., 2021; Luca et al., 2021; Siła-Nowicka et al., 2016). These applications include real-time navigation, such as Google Maps and Apple Maps, as well as social media platforms, such as Twitter (now referred to as X), Instagram, and Facebook. These datasets have offered opportunities to advance the understandings of how individuals travel and interact with their surrounding environments (Jiang, Huang, et al., 2021; Rahman et al., 2016; J. Yang et al., 2019; Yuan et al., 2021).

Mobile location datasets come in various forms, including the traditional call detailed records (CDRs), location information retrieved from an application, and commercially curated datasets (Hsu et al., 2024; Isaacman et al., 2012; Z. Li et al., 2024; K. Zhao et al., 2016). Traditional CDRs capture cell-to-tower communications and record such activities. Although CDRs record cell users' movements across the space, it lacks more background context, such as the demographic characteristics of the cellphone users and the motivation behind the movements (Ranjan et al., 2012, p. 20; Wesolowski et al., 2013; Z. Zhao et al., 2016). On the other hand, location information retrieved from installed applications, such as Twitter, provides more context, but at the same time, it suffers from representative biases and limited access to the complete dataset (Jiang et al., 2019; Malik et al., 2015; Mislove et al., 2011).

Some commercial data companies collect, organize, and provide mobile location data through partnerships with multiple applications, aiming to provide a more comprehensive and well-represented mobility dataset. These companies include SafeGraph (now as Advan Patterns), Cuebiq (now called Spectus), and X-mode, to name a few. These companies provide anonymized movement patterns at an aggregated spatial unit, providing valuable mobility information while protecting privacy at the same time. However, the exact data collection and sampling processes remain unclear, introducing possibilities of biases in the datasets (Coston et al., 2021; Hsu et al., 2024; Z. Li et al., 2024; Pepe et al., 2020).

Besides mobile location data, location information retrieved from GPS-enabled vehicles is also an important data source (He et al., 2020; K. Zhao et al., 2016; Zheng et al., 2008). Such data is usually retrieved from taxis, ride-sharing services, or public transit, where tracking vehicles' locations is essential for their services. These datasets offer a real-time or near real-time tracking of vehicles, based on which researchers can gain insightful understandings of the urban dynamics, such as traffic flows, road conditions, and human movements (Ceder, 2021; Riascos & Mateos, 2020; Siangsuebchart et al., 2021; A. Wang et al., 2020). Under the open data policies in some cities, these datasets are granted with full access, without sampling or truncating, providing a more comprehensive view of mobility (Jiang, Huang, et al., 2021; Mukherjee & Jain, 2022; Z. Wang et al., 2024; Willis & Tranos, 2021). These datasets are utilized in various applications, including traffic management (Dokuz, 2022; C. Liu et al., 2021; Takahashi et al., 2004), urban event detection (Jiang et al., 2022; Kuang et al., 2015; Zhu & Guo, 2017), route optimization (Chuah et al., 2016; Lin et al., 2012; Qu et al., 2019), and urban planning (H. Li et al., 2021; Y. Liu et al., 2012). However, the issue of representativeness also exists as these datasets only capture the mobility patterns from individuals preferring a specific type of transportation mode (Anda et al., 2017; Buehler, 2011; Convery & Williams, 2019; Ricord & Wang, 2023).

Since the Covid-19 pandemic, there has been a growing focus on human mobility and its association with infectious diseases. Human mobility datasets have emerged as essentials in related scientific research (Benita, 2021; Hu et al., 2021; Lessani et al., 2023). To promote rapid research and benefit society, certain data providers granted open access to researchers. Notably, SafeGraph stood out as a key contributor. Their mobility dataset, the Social Distancing Metric, was released for free scientific research use during the early stage of the pandemic, until April 2021 when the dataset stopped updating and stopped free access (Chang et al., 2021; Y. Kang et al., 2020; Lu & Giuliano, 2023). The SafeGraph data includes information on movements and points-of-interests at the census block group (CBG) level, providing researchers with a rich source of information to explore human movements. SafeGraph, along with other commercially curated mobility datasets, has significantly contributed to the data-driven research in multiple disciplines, including infectious diseases (Huang, Li, Lu, et al., 2020; Z. Li et al., 2020; Ning et al., 2023; Susswein et al., 2023; Weill et al., 2020), extreme weather responses (P. Chen et al., 2023; Huang et al., 2024; Vachuska, 2024), disaster management (Dargin et al., 2021; Jiang et al., 2023; Z. Wei & Mukherjee, 2023), and urban planning (Jay et al., 2022; Juhász & Hochmair, 2020). To help understand the population representativeness in this widely used mobility dataset, SafeGraph provides an assessment of sampling biases in their datasets (Squire, 2019). In addition, a few studies have investigated the sampling rate of SafeGraph datasets by comparing the sampled device amount with census data or survey information (Coston et al., 2021; Z. Li et al., 2024; Liang et al., 2021; J. Wang et al., 2021).

Numerous studies have utilized mobility data from GPS-enabled vehicles and SafeGraph datasets to analyze human movement patterns (e.g., Colombo & Borri, 2020; Gupta et al., 2020; Handley et al., 2021; Liang et al., 2018; Ma et al., 2015; Parker et al., 2021; Santi et al., 2014; Szymanski & Malinowski, 2020; Tang et al., 2015; D. Zhang et al., 2014; K. Zhao et al., 2016; B. Zhang et al., 2020; L. Zhang et al., 2020). Despite this, there has been limited research that directly compares these two data sources to assess how comprehensively each one captures and reflects human mobility behaviors. This gap is significant because GPS-enabled vehicles and SafeGraph datasets differ in their data collection methods and are subject to distinct biases, which may influence how accurately they represent movement patterns. Understanding these differences is crucial for ensuring the reliability of mobility data across research and policy applications. To bridge this gap, our study presents a comparative analysis of origin-destination travel patterns using SafeGraph and taxi datasets across various neighborhoods in New York City. By examining the relationships between these movement flows and neighborhood characteristics, we aim to provide insights into the representativeness and limitations of these datasets. Specifically, this study seeks to answer the following research questions:

- What are the predominant travel flow patterns identified in taxi and SafeGraph data across diverse neighborhoods in NYC?
- How do the socioeconomic and demographic characteristics of neighborhoods relate to the flow patterns identified above?
- What similarities and differences exist in travel patterns among different neighborhoods of NYC when comparing taxi and SafeGraph datasets?
- Based on the analysis of these findings, what insights can be drawn regarding the ability of each dataset to capture and reflect various aspects of human mobility within populations?

The primary procedures of the study proceed as follows. Initially, we perform clustering analysis using demographic, socioeconomic, and commuting behavior variables to define distinct neighborhood boundaries in NYC and explore the unique population characteristics of each neighborhood. Subsequently, we summarize the travel flows captured by both the SafeGraph and taxi datasets between each neighborhood. Finally, through a comparison of the recorded travel flows from these two datasets, we pinpoint any under-represented trip flows and analyze transportation mode preferences in particular neighborhood characterized by varying population demographics.

The remainder of this paper is organized as follows. Section 2 reviews existing studies, providing background information in taxi data, mobile location data, and their applications in human mobility studies. Section 3 introduces datasets and the study area. Section 4 describes the methodology used in this study. Section 5 presents the results and discussions about the findings. Finally, section 6 discusses the limitations and concludes this study.

## 2. Background

### 2.1. Taxi Data as a Window into Urban Mobility

Mobility information extracted from GPS-enabled taxis has emerged as a valuable data source for understanding human mobility in urban environments (J. Tang et al., 2015; D. Zhang et al., 2014; K. Zhao et al., 2016). Taxi datasets provide rich insights, enabling researchers to analyze travel patterns, traffic flow, and urban dynamics. Lyu et al. (2021) offers a comprehensive review of taxi and ride-sharing services.

Taxi datasets generally fall into two categories. The first type is origin-destination (OD) travel records, which include only the origin and destination of each trip, without any information on the routes taken (Jiang, Guo, et al., 2021). The second type consists of floating point data, where each vehicle transmits its location to a database at regular intervals (H. Chen et al., 2021; Kong et al., 2016; Shen et al., 2017). This data allows for the reconstruction of taxi routes; however, it typically lacks information on the taxi's status, such as whether it is occupied or empty. In this section, we focus on the first type, OD travel records, where only the origin and destination locations of each trip are recorded, and route information is not available.

*Research Directions in Taxi Data Analysis*
Existing studies using OD travel records fall into three key areas:

1. *Urban Dynamics and Travel Pattern Analysis*: Researchers use clustering and classification techniques to identify travel hotspots and common patterns. These insights help city planners understand how urban spaces are used and assist businesses with marketing strategies (Scholz & Lu, 2014; Yao et al., 2018; Zhu & Guo, 2014).
2. *Human-Environment Interactions*: This line of research investigates the intricate relationship between taxi demand and various land uses, such as commercial and residential areas, alongside social determinants like income and education levels, commuting patterns, and public transportation accessibility, providing insight into how individuals interact with their human-built environment. These insights are invaluable for urban planners, offering data-driven guidance to optimize transportation systems and enhance urban livability (Peng et al., 2012; Choi et al., 2022; Ghaffar et al., 2020; Pan et al., 2012; Qian et al., 2015; Yang et al., 2018; Yu et al, 2021).
3. *Event Detection and Anomaly Identification*: Leveraging large-scale taxi datasets, researchers have detected urban events (e.g., parades, policy changes) and monitored road conditions using statistical and machine learning approaches (Markou et al, 2017; Jiang et al., 2022; Zhu & Guo, 2017; Zhang et al., 2015).

*Limitations in Taxi Data*
A major limitation of taxi-based studies is the representativeness of the data. Since a significant portion of the population does not rely on taxis for their daily travel, taxi records may exclude a substantial demographic, raising concerns about their ability to accurately reflect the diversity of urban populations. As a result, researchers have explored various factors that influence transportation mode choices. One key factor is access to public transit—studies consistently show a negative correlation between the availability of efficient public transportation and taxi usage (Gebeyehu & Takano, 2007; Qian & Ukkusuri, 2015; J. Tang et al., 2019; Tu et al., 2018; Z. Yang et al., 2018). In other words, people tend to use taxis less in areas where public transit is easily accessible. The built environment is also a critical determinant: commercial zones with corporate offices and shopping centers are strongly associated with higher taxi use, while residential areas show mixed results (Bi et al., 2020; B. Li et al., 2019; T. Li et al., 2019; L. Zhang et al., 2020; X. Zhang et al., 2019). Furthermore, external factors like weather conditions and social networks can significantly influence individual decisions regarding taxi usage (Böcker et al., 2016; Pike & Lubell, 2016).

## 2.2. Cell Phone-Based Location Data: Strengths and Challenges

Cell phone-based location data has gained prominence as an important tool for understanding human mobility patterns (e.g., Colombo & Borri, 2020; Gupta et al., 2020; Handley et al., 2021; Parker et al., 2021; D. Zhang et al., 2014). Companies like SafeGraph, Cuebiq, and X-mode collect location data through partnerships with mobile apps. These datasets offer extensive coverage, allowing researchers to examine large-scale mobility trends efficiently (Y. Kang et al., 2020).

*Advantages and Limitations*
The primary advantage of cell phone-based location data is its accessibility and cost-effectiveness for researchers. By eliminating the need for labor-intensive data collection, this data provides scientists with broad, comprehensive geospatial insights into individual movements (Y. Kang et al., 2020). However, significant challenges persist, particularly in validating the data against actual real-world movements. Since these datasets are anonymized and aggregated to protect privacy (Hu et al., 2021; Lu & Giuliano, 2023), the data collection methods often remain opaque, raising concerns about its reliability and accuracy (Li et al., 2024). Moreover, the proprietary nature of these datasets restricts transparency, as researchers may not have access to the methodologies used to estimate locations, such as determining users' home locations. This lack of openness complicates efforts to scrutinize or reproduce findings, potentially undermining confidence in the results (Wang et al., 2021).

*Representativeness and Sampling Bias in SafeGraph Data*
SafeGraph has become a widely utilized proxy for human mobility and population distribution, but concerns about representativeness and sampling biases persist. Wang et al. (2021) compared SafeGraph data with Census data in North Carolina, finding that sampling rates were balanced across demographic groups. However, Coston et al. (2021) observed that SafeGraph data had lower sampling rates for older individuals and non-white populations during the 2018 general election in North Carolina. Liang et al. (2021) assessed the validity of mobile device data for monitoring national park visitors by comparing it with traditional surveys and counts. They identified discrepancies in visitor demographics, particularly education and income levels, due to platform biases and the exclusion of international visitors. These differences highlight the limitations of mobile data in accurately representing the full range of park visitors compared to traditional methods.

**2.3 Gaps in Human Mobility Research Using GPS-enabled Taxi and SafeGraph Data**

As mentioned in the previous section, while GPS-enabled Taxi and SafeGraph datasets are commonly used to study human mobility, both face significant challenges regarding representativeness. However, few studies have directly compared these datasets to understand the distinct mobility patterns each capture. Such a comparison is essential because differences in data collection methods and inherent biases affect their accuracy in reflecting human movement. Specifically, analyzing the differences in mobility patterns that each dataset captures across various neighborhoods—such as pedestrian-heavy, public transit-dominated, and car-dependent areas—along with key socioeconomic factors like car ownership rates, transportation modes, and commuting times, would offer valuable insights for more accurate urban mobility analysis. Understanding these distinctions is vital for studies on urban planning, transportation, and addressing socioeconomic disparities in mobility access.

**3.   Datasets and Study Area**
**3.1. Study Area**

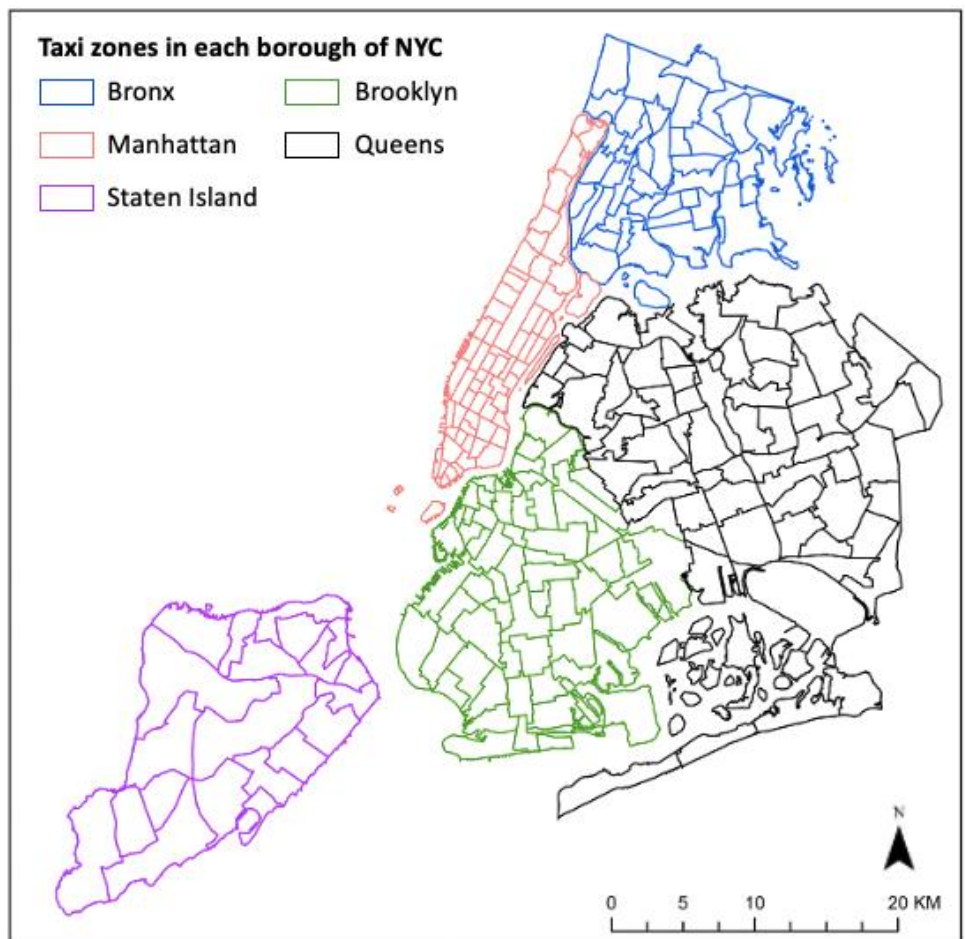

Figure 1. Study area – New York City (NYC)

Our study focuses on New York City (NYC), where we collected both taxi and SafeGraph data generated within the city's boundaries. Figure 1 shows a map of NYC. NYC consists of five boroughs, and each borough is a county in the state: Brooklyn (Kings County), Bronx (Bronx County), Manhattan (New York County), Queens (Queens County), and Staten Island (Richmond County). The population of NYC is more than 8.8 million based on an estimation by Census Bureau (U.S. Census Bureau, n.d.). It has the highest population density of 29,302 people per square mile. Due to the high density and expensive land price, car ownership is low in NYC, especially in Manhattan. Public transit (subways and buses), bike-sharing, ridesharing, and taxis play important roles in the daily travel for people living in NYC.

### 3.2. Datasets
#### 3.2.1. Taxi data

The New York City Taxi and Limousine Commission (NYC TLC) provide the taxi datasets under an open data policy (NYC TLC, n.d.). NYC TLC manages the operation and licensing of taxicabs and ride-sharing services within the city. This dataset covers a wide range of vendors, from the traditional taxicabs to major ride-sharing applications, such as Uber and Lyft, as well as various local ride-sharing service providers.

NYC TLC employs a zoning system to record the origin and destination of each taxi trip. It consists of 263 zones, with 262 covering five boroughs of NYC and an additional zone designated for Newark Airport in New Jersey. According to NYC TLC, the taxi zones roughly follow the boundary of neighborhoods (*TLC Trip Records User Guide*, 2019). Each trip record covers the following information: vendor ID, pick-up date time, drop-off date time, passenger count, trip distance, rate code, pick-up location ID, drop-off location ID, payment type, fare amount, extra charge, MTA tax, tip amount, tolls amount, improvement surcharge, total amount, congestion surcharge.

In this study, we downloaded trip records from the NYC TLC website and summarized travel records from various vendors, including the traditional taxicabs and multiple ride-sharing applications. To be consistent with SafeGraph data, the NYC taxi data used in this study spans from January 1st, 2019, to March 31st, 2021. During the data processing step, we temporally aggregated taxi trip records to a daily unit. The spatial aggregation level remains as the taxi zone.

#### 3.2.2. SafeGraph data

Mobile location datasets are obtained from SafeGraph, a company that specializes in providing geospatial data and analytics. During the Covid-19 pandemic, SafeGraph made its mobility related datasets accessible with no charge for scientific research. However, in early 2022, SafeGraph started charging their datasets. The temporal range of SafeGraph mobility dataset used in this study is between January 1st, 2019, to March 31st, 2021. These datasets were downloaded when they were available at no cost.

SafeGraph utilizes a panel of anonymous mobile devices to derive aggregated home locations. The determination of home location is based on nighttime locations recorded by these devices over a span of six weeks. For privacy concerns, the home location of each device is aggregated to the census block group (CBG) level (SafeGraph, 2020). The SafeGraph movement flows were organized from their Social Distancing Metrics dataset. In this dataset, each CBG is considered as an origin CBG and has an entry in the dataset. In the column for destination CBGs, it records two pieces of information: identification of the destination CBG and the number of devices that stopped in the given CBG for more than one minute.

In this study, we applied a spatial filter to so that only trips occurred between two CBGs in NYC were retrieved. These NYC SafeGraph trips were then spatially aggregated to the taxi zone level (see section 4.1 for more detail). The temporal unit for the SafeGraph data remained as daily.

### 3.2.3. ACS data

Demographic and socio-economic variables are available at census tract level from American Community Survey (ACS). We used 2017-2021 estimation data to extract relevant features. These datasets were acquired from the National Historical Geographic Information System (NHGIS), maintained by the Minnesota Population Center at the University of Minnesota. NHGIS organizes and provides easy access to census data, ranging from population, economic, housing, and commuting (Manson et al., 2023).

## 4. Method

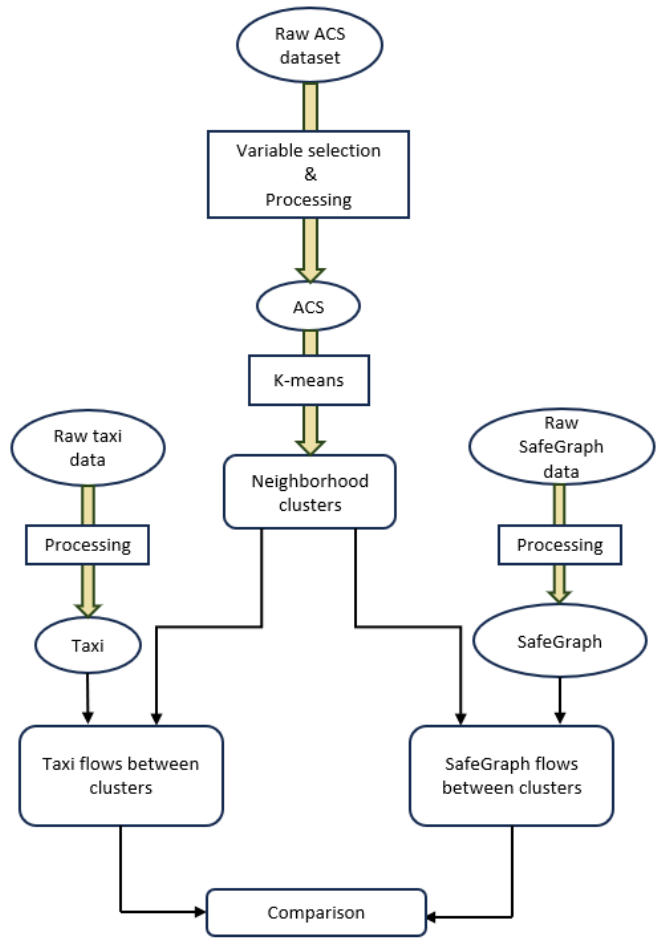

Figure 2. Workflow for this study.

This study comprises four main steps, as shown in Figure 2. The first step is data processing and preparation. In this study, we utilize taxi zone as the spatial unit. Therefore, all the variables are aggregated and summarized at the taxi zone level. The second step is clustering analysis. We conduct clustering analysis using demographic, socioeconomic, and commuting behavior variables to delineate neighborhood boundaries in NYC. Each neighborhood cluster has its unique characteristics, which we identify and explain for each neighborhood. Subsequently, we summarize trip flows based on taxi trip records and SafeGraph dataset between each neighborhood cluster. Taking each cluster as origin and destination, we generate the OD flow chord diagrams for both taxi and SafeGraph flows. Lastly, we identified under-represented trip flows by comparing the OD flows between taxi and SafeGraph datasets.

## 4.1. Data processing

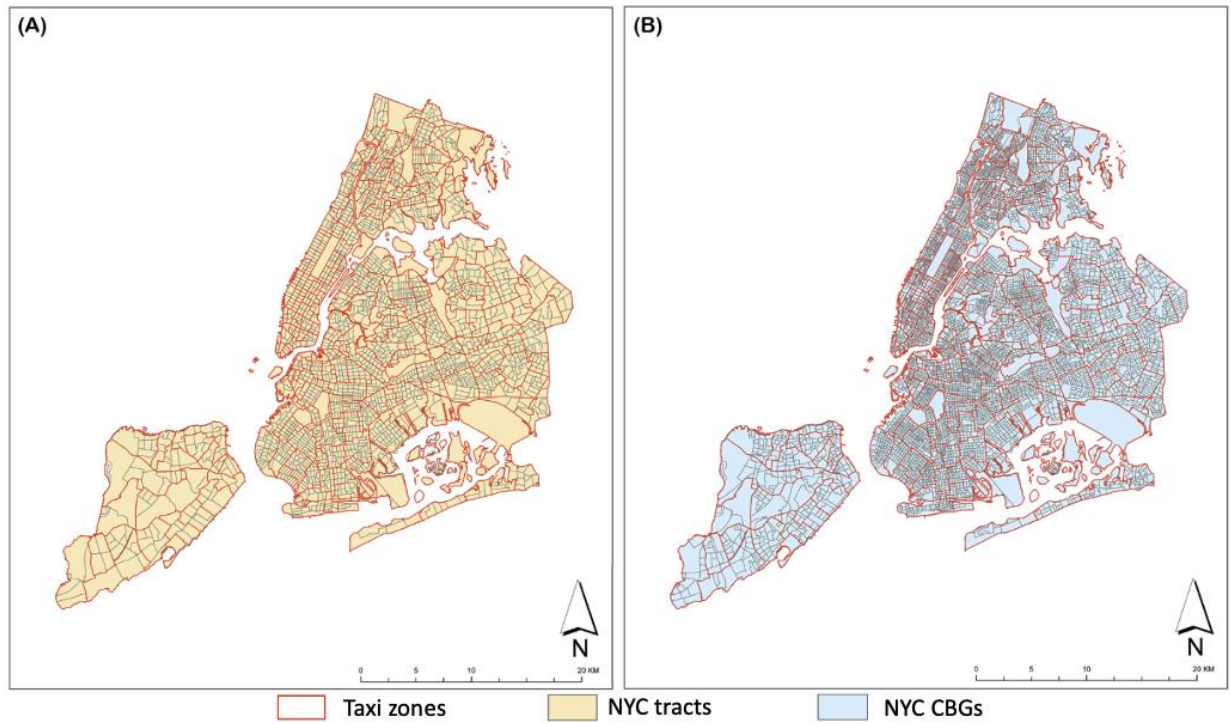

Figure 3. Comparison between taxi zone boundary with (A) tract boundary and (B) CBG boundary.

In this study, we use taxi zones as the spatial unit. The taxi zone partitions mostly follow the neighborhood divisions in NYC. Figure 3A demonstrates the comparison between taxi zones and census tracts and Figure 3B demonstrates the comparison between taxi zones and the census block group. We created a matching table to aggregate census tracts and CBGs into taxi zones. All the human mobility flows and ACS variables are processed to the taxi zone level. SafeGraph mobility dataset records travel flows between CBGs. We also re-calculated SafeGraph mobility flows into taxi zone to taxi zone based on the which taxi zone each CBG belongs to.

**4.2. Neighborhood clustering analysis**

We created neighborhood boundaries based on similarities in demographic, socioeconomic, and commuting behavior. Table 1 lists specific variables used for clustering neigborhood in this study. The variables representing these characteristics were initially aggregated at the census tract level, downloaded from NHGIS, and then re-aggregated to align with taxi zone boundaries. The specific variables were selected based on existing literature on neighborhood analysis (Delmelle, 2017; Foote & Walter, 2017; Han et al., 2023; F. Wei & Knox, 2014), where researchers have used K-means clustering to delineate neighborhood using demographic and socioeconomic variables. Previous studies typically included variables such as the proportion of young adults and seniors, education levels, income, race, and ethnicity to delineate neighborhood boundaries. In addition to these commonly used factors, we have incorporated transportation-related variables, such as commuting time, means of transportation, and work-from-home status, sourced from the American Community Survey, as this study focuses on transportation behavior.

Table 1. Selected demographic, socioeconomic and commuting behavior variables used in neighborhood clustering analysis.

| Variable Name | Description |
| --- | --- |
| malePercent | Percentage of male in the total population |
| age<18Percent | Percentage of population that is 18 years or younger |
| age>60Percent | Percentage of population that is 60 years or older |
| collegeAbove | Percentage of population has college degree or above |
| poverty | percentage of population whose income is below poverty level in the past 12 months |
| unemployment | Percentage of population 16 years and over who was unemployment |
| renterPercent | Percentage of housing units that are renter occupied |
| noCarPercent | Percentage of households without a car |
| medianIncome | Median household income in the past 12 month (in 2021 inflation-adjusted dollars) |
| noMove | Percentage of population lived in the same house one year ago |
| housingCostPerc | Housing cost as a percentage of household income in the past 12 months |
| comDriveAlone | Percentage of population whose means of transportation to work is drive alone |
| comCarpool | Percentage of population whose means of transportation to work is carpool in 2 or more persons |
| comPublicTransit | Percentage of population whose means of transportation to work is public transportation (excluding taxicab) |
| comWFH | Percentage of population whose means of transportation to work is work from home |
| comTaxi | Percentage of population whose means of transportation to work is taxicab |
| comMotorcycle | Percentage of population whose means of transportation to work is motorcycle |
| comBikeWalk | Percentage of population whose means of transportation to work is walk or bicycle |
| commute<10min | Percentage of population whose travel time to work is less than 10 minutes |
| commute10-29min | Percentage of population whose travel time to work is 10 to 29 minutes |
| commute30-59min | Percentage of population whose travel time to work is 30 to 59 minutes |
| commute>60min | Percentage of population whose travel time to work is more than 60 minutes |
| White | Percentage of white population |
| Black | Percentage of black population |
| Asian | Percentage of Asian population |
| Hispanic | Percentage of Hispanic/Latino population |
| foreignBorn | Percentage of foreign-born population |

Neighborhood boundary delineation method is commonly used in neighborhood analysis, which usually involves clustering algorithms that groups neighborhoods with similar characteristics together into one cluster (Delmelle, 2017; S. Rey et al., 2018; S. J. Rey et al., 2011; R. Wei et al., 2021). The purpose of doing so is to reduce the dimensionality of variables and identify common characteristics for neighborhoods. Various clustering algorithms have been developed for this purpose. In this study, we used k-means clustering method, which is one of the most commonly used clustering methods for neighborhood boundary delineation (Delmelle, 2016; Han et al., 2023; W. Kang et al., 2020; Y. Kang et al., 2020; Ling & Delmelle, 2016; F. Wei & Knox, 2014). It has been used in multiple studies to identify neighborhood clusters. The optimal value of k (i.e. the number of clusters) is determined by the elbow method (Kodinariya & Makwana, 2013).

We applied the K-means clustering method to taxi zones where the variables listed in Table 1 were available. Some zones lacked these variables, so we aggregated them and labeled them as 'Undefined' zone. In this study, the optimal number of clusters was determined to be five, resulting in five distinct clusters.

### 4.3. Metrics for pairwise OD matrices comparison

We organized both taxi and SafeGraph datasets into OD travel matrices. We summarized the inter-cluster travel flows recorded in taxi and SafeGraph datasets, according to the clustering analysis results from the previous step. In this step, we generated a taxi OD travel matrix and a SafeGraph travel matrix. In both matrices, each row records trips starting from the corresponding cluster and each column records trips ending in the corresponding cluster. We can consider two ways to compare frequency in pairwise comparison. On the one hand, comparison of absolute frequency between two datasets is helpful to understand and measure the difference between the two datasets in terms of the number of collected records and data availability. On the other hand, comparison of relative frequency between the two datasets is valuable to identify bias of each dataset. In this study, we focus not only on absolute frequency but also on relative frequency. As the difference of the total numbers of trips between two datasets in the study area is not significant (536M vs. 602M), interpretation of absolute frequency comparison is similar to that of relative frequency comparison.

To clarify our metrics for measuring relative frequency, we define notations. Let $E_{ij}(G)$ be a set of trips starting from zone $i$ and ending in zone $j$ in dataset G. Let $|E|$ be the cardinality of set $E$. $|E_{ij}(G)|$ denotes the number of trips from $i$ to $j$ in G. To compare two datasets which have different cardinality, we first normalize the frequency of OD trips. Let $RF\left(E_{ij}(G)\right) = \frac{|E_{ij}(G)|}{|E(G)|}$ be the relative frequency of $E_{ij}(G)$, where $|E(G)| = \sum_i \sum_j |E_{ij}(G)|$ is the total number of trips in G,

$0 \leq RF\left(E_{ij}(G)\right) \leq 1$, and $\sum_i \sum_j RF\left(E_{ij}(G)\right) = 1$. To clarify, relative frequency refers to proportion or percentage of times a trip occurs relative to the total number of trips. To measure relative difference of trip between two datasets, we define relative frequency ratio or RFR as follows:

$$RFR(E_{ij}, G_a, G_b) = \frac{RF\left(E_{ij}(G_a)\right)}{RF\left(E_{ij}(G_b)\right)}, \quad \text{Eq. 1.}$$

where $G_a$ and $G_b$ are datasets. This ratio provides a quantitative comparison. If $RFR(E_{ij}, G_a, G_b) > 1$, it means that relatively more frequent trips from $i$ to $j$ in $G_a$ are observed than that in $G_b$. In this case, we can conjecture that the trips in $G_a$ is over-sampled or trips in $G_b$ is under-sampled. If RFR is larger, the conjecture is more convincing. If methods collecting data and sampling for two datasets are the same or similar, we can expect that each RFR is close to 1.

One downside of RFR is that it is difficult to intuitively compare numerical values between two ranges, i.e., $RFR < 1$ and $RFR > 1$, because of the scale difference. For that reason, we define the logarithm of relative frequency ratio (LRFR) as follows:

$$\begin{aligned} LRFR(E_{ij}, G_a, G_b) &= \log_2 RFR(E_{ij}, G_a, G_b) \\ &= \log_2 \frac{RF\left(E_{ij}(G_a)\right)}{RF\left(E_{ij}(G_b)\right)} \\ &= \log_2 RF\left(E_{ij}(G_a)\right) - \log_2 RF\left(E_{ij}(G_b)\right). \quad \text{Eq. 2.} \end{aligned}$$

If $LRFR(E_{ij}, G_a, G_b) = 0$, it means that two datasets $G_a$ and $G_b$ have the same relative frequency of trips from $i$ to $j$. $LRFR(E_{ij}, G_a, G_b) = 1$ means the relative frequency of trips from $i$ to $j$ in $G_a$ are two times (i.e., 2 to the power of 1) higher than those in $G_b$. Conversely, $LRFR(E_{kl}, G_a, G_b) = -1$ means the relative frequency of trips from $l$ to $k$ in $G_b$ are two times (i.e., 2 to the power of 1) higher than those in $G_a$. That is, absolute values of LRFR are symmetric, and it is easier to comprehend difference of values of LRFR than those of RFR. Since we do not expect exponentially huge difference of relative frequency between two datasets, a smaller base of logarithm is better to interpret. Thus, this study uses a binary logarithm.

## 5. Results and discussion
5.1. Neighborhood clustering

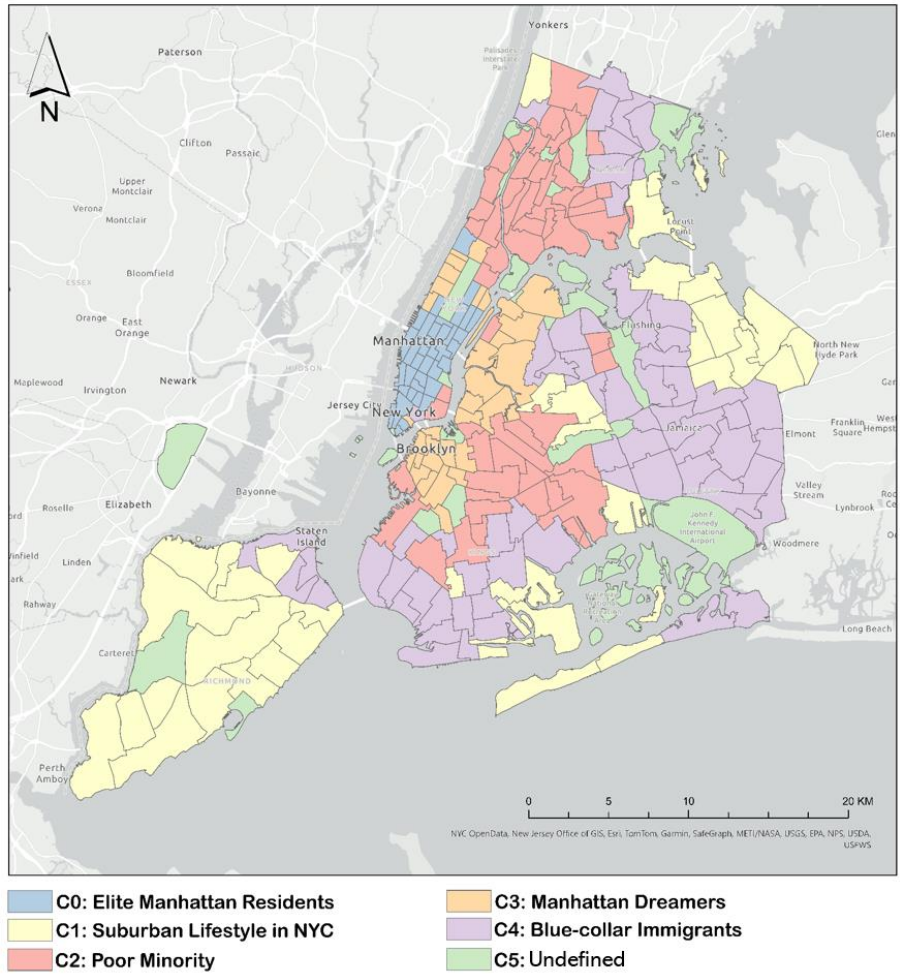

Figure 4. Neighborhood clustering results. Taxi zones are categorized into 6 clusters based on their demographic, socioeconomic, and community characteristics.

Figure 4 shows the map of neighborhood clustering results. In this map, there are 5 different clusters generated based on the demographic, socioeconomic, and commuting behaviors. It is important to note that these variables are only available for zones with registered residents. In NYC, there are also several zones without enough registered residents, such as Central Park, JFK airport, and other open spaces. We labeled these zones as our 6th cluster with "undefined zones" in the later results and discussions. Figure 5 presents the characteristics of each cluster using Z-score means, where a Z-score reflects how many standard deviations a data point is from the mean. A negative Z-score indicates that the data point is below the mean. In addition to the insights provided by Figure 5, which highlights variations in Z-score means for each variable, the distribution of all variables listed in Table 1 is visualized through box plots, further illustrating the detailed variation of each variable across the five identified clusters (neighborhoods). We also labeled each cluster based on its defining characteristics, derived from the Z-score means shown in Figure 5, as well as their geographical locations. The detailed descriptions of each cluster are found below:

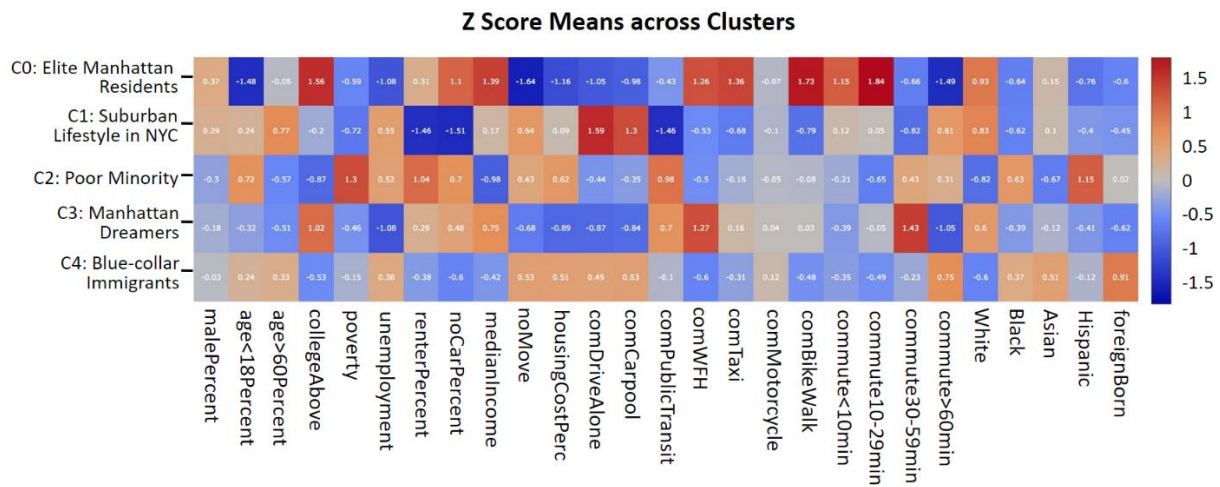

Figure 5. Z-score means for all the variables at each cluster.

C0 Elite Manhattan residents: these zones are mostly in the lower and midtown Manhattan. People living in C0 have high income, high percentage with college or above education. Low percentages of poverty and unemployment. At the same time, the percentage of households without car is high. Regarding their commuting behavior, taxi, bike, or walk are the three main commute modes for people living here. In addition, they have short commute time, usually less than 30 minutes. People living in this part are likely those who also work for one of the companies in Manhattan with high salary, and thus can afford to live in Manhattan. This area has the highest percentage of residents who have relocated within the past year, indicating a significant number of recent movers. Demographically, the percentage of white population is highest among all zones.

C1 Suburban lifestyle in NYC: these zones are located on the outskirts of NYC. Most of Staten Island belongs to this cluster. Also, the northeast part of Queens neighborhoods also belongs to this cluster. Compared to Manhattan and its immediate surrounding areas, these outskirt neighborhoods have a lower population density. More single house residential areas are found. Most people living here are homeowners with private vehicles. Therefore, the main commute method is to drive, either drive alone or carpool. Commuting by public transit, taxi, bike or walk percentage is low. The elderly population is higher. Demographically, the white population is significantly above average, marking the second highest among all neighborhoods.

C2 Poor minority: the majority of these neighborhoods are located in Bronx and central Brooklyn. These neighborhoods are featured with high poverty and unemployment rates. Renter percentage is high. Their median income is the lowest among the 5 clusters. In addition, the housing costs as a percentage of household income in past 12 months are highest across all neighborhoods, indicating that a significant portion of their income was allocated to housing expenses. The household with no car percentage is high, indicating that a large percentage of households within this zone do not have a vehicle. Therefore, public transit is the main commute method. The percentage with more than 30 minutes commute time is higher than average, but not the highest among these groups. Demographically, black and Hispanic populations are high.

C3 Manhattan (elite) dreamers: these neighborhoods are located in Brooklyn and Queens, right across Hudson River from Manhattan. Also, it includes some neighborhoods in the Upper East Side in Manhattan. These cluster shows similar patterns with C0 in terms of socioeconomic and demographic composition. It has the second highest college education percentage and second highest median income; both are only behind C0 neighborhoods. Like C0, poverty and unemployment are low. The main difference from C0 is commuting. The main commute method for C3 neighborhoods is by public transit as indicated in Figure 6 and the commuting time is mostly between 30 to 59 minutes. Demographically the white population is high.

C4 Blue-collar immigrants: these neighborhoods are located in the majority of Queens, south Brooklyn, northern tip of Staten Island, and north Bronx. The most significant characteristic of these neighborhoods is high foreign-born percentage, with higher than average Asian and black population. It has the second lowest median income among these 5 clusters and also the second lowest college education percentage. Housing cost is high. The percentage of households without a car is the second lowest. The percentage of commuting by driving is higher than average. In addition, the percentage of population whose commute time is more than 60 minutes is the highest in this cluster.

C5 Undefined: Unlike the previous five clusters, this zone lacks American Community Survey (ACS) data on socioeconomic, demographic, and commuting characteristics due to little or no residential population. As a result, it was excluded from the K-means clustering and is considered an undefined neighborhood in the analysis. Since ACS data estimate nighttime populations, and no population exists in these areas overnight, variables like race or income cannot be measured. However, these areas may experience significant daytime population flow, primarily consisting of parks (e.g., Central Park) and airports (JFK and LaGuardia).

As examining the transportation mode offers crucial insights into the notable disparity between taxi trips and SafeGraph trips, and vice versa, we examined the proportion of transportation modes within each neighborhood cluster. While Figure 5 allows comparison of the extent to which the percentage of the population relies on different modes of transportation across clusters, it falls short in revealing the proportion of each mode within individual neighborhoods. To address this gap, we created a percent stacked column chart to provide a comprehensive view of transportation mode distribution within each neighborhood. The findings from this chart will be used to explain why taxi trips are significantly higher than SafeGraph trips, and vice versa in the subsequent sections.

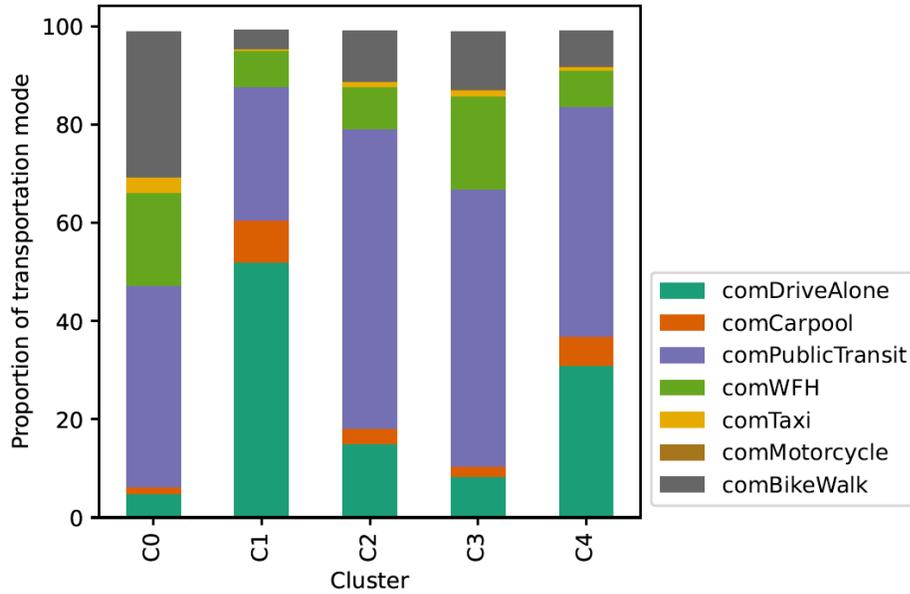

Figure 6. Proportion of transportation mode in each neighborhood (C0: Elite Manhattan Residents, C1: Suburban Lifestyle in NYC, C2: Poor Minority, C3: Manhattan Dreamers, C4: Blue-collar Immigrants, C5: Undefined). Table 1 provides a full description of short labels in the legend.

## 5.2. Comparing taxi and SafeGraph datasets
### 5.2.1. Taxi flows

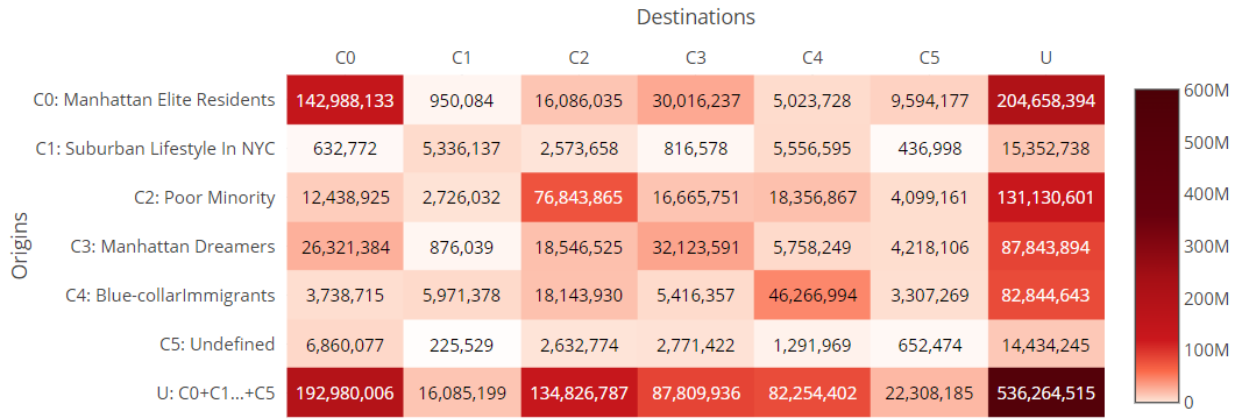

Figure 7. Taxi trip flows between clusters. U represents all clusters. For example, the cell in the bottom left corner represents the number of trips starting from any cluster and ending in C0. Similarly, the cell in the top right corner represents the number of trips starting from C0 and ending in any cluster.

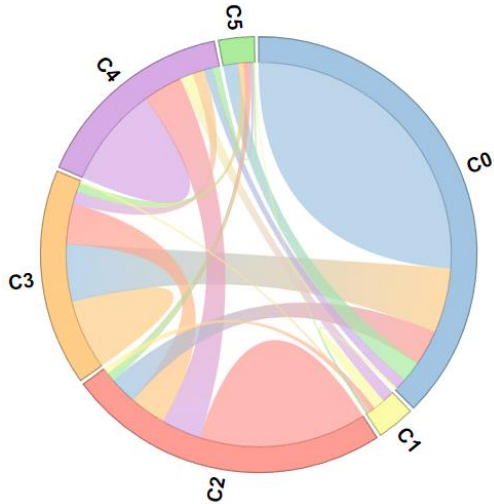

Figure 8. Chord diagram presenting the number of taxi trips between clusters. C0: Elite Manhattan Residents; C1: Suburban Lifestyle in NYC; C2: Poor Minority; C3: Manhattan Dreamers; C4: Blue-collar Immigrants; C5: Undefined.

Figure 7 presents the origin-destination taxi trip flow matrix, while Figure 8 offers an illustrative chord diagram for the flows, offering a direct quantitative comparison of trip volumes across each cluster-to-cluster traveling. Over the 27-month period, a total of 536,264,515 taxi trips were recorded by NYC TLC. This includes trips conducted with both traditional taxicabs and ride-sharing services.

Notably, 142,988,133 trips, consisting of about 26.67% of the total trips, occurred inside C0 (Elite Manhattan residents). C0 zones are characterized by a large number of world-class companies and historically iconic landmarks, such as Wall Street, New York Stock Exchange, Times Square, Rockefeller Center, and the Empire State Building. Their status as the global financial and business centers significantly boosted land prices, making it among the most expensive land worldwide. These Elite Manhattan Residents primarily engaged in high-paying occupations within the world-class companies nearby. Remarkably, a large percentage of C0 (Elite Manhattan residents) residents rely extensively on taxi services for their daily commuting and other travel demands, as their car ownership is very low. This results in the highest volume of C0-to-C0 taxi trip flows. Furthermore, the concentration of businesses within C0 (Elite Manhattan residents) also contributes to a significant portion of these C0-to-C0 taxi trips, as these upscaled companies also rely on taxi or ride-sharing services for their business trips. Other than intra-C0 travels, 9,594,177 trips starting from a C0 (Elite Manhattan residents) ended in a C5 (Undefined zone), consisting of 4.7% of all the taxi trips started from C0. Most of these trips were heading to an airport. This pattern emphasizes the crucial role of taxis in addressing the mobility demands of C0 (Elite Manhattan residents). Moreover, another 30,016,237 trips originating from a C0 (Elite Manhattan residents) ended in a C3 (Manhattan Dreamers), which is about 14.7% of all the trips starting from a C0 zone. Compared to C0 (Elite Manhattan residents), C3 (Manhattan Dreamers) have lower land prices, which makes it more affordable for living. On the other hand, C3 (Manhattan Dreamers) residents are more likely to endure longer commute times to downtown or midtown Manhattan.

The second largest taxi trip flow was within C2 (poor minority), which is about 14.33% of all the taxi trips. Additionally, among the taxi trips starting from each cluster (i.e., cells in the right-most column excluding the bottom cell in Figure 7), those starting from C2 show the second highest number of trips. Similarly, among the taxi trips ending in each cluster (i.e., cells in the bottom row excluding the right-most cell Figure 7), trips ending in C2 demonstrate the second highest frequency.

When considering the population residing in C2 (poor minority) areas, the second-highest frequency of taxi trips within, starting from, and ending in this zone mentioned above can be attributed to the relatively large population residing in this area – i.e., C2 experiences significant demand for taxis, largely driven by its dense population. Figures 9 and 10 highlight this finding. Figure 9 illustrates the number of taxi trips: (A) within each cluster, (B) starting from each cluster, and (C) ending in each cluster. In contrast, Figure 10 displays normalized taxi trips per population: (A) within each cluster, (B) starting from each cluster, and (C) ending in each cluster. In all three charts of Figure 9, C2 shows the second-highest number of taxi trips. However, upon examination of the normalized taxi trips per population in Figure 10, C2 no longer holds the second position, and there is less disparity between C2 and C3/C4 across all categories (A), (B), and (C) compared to Figure 9.

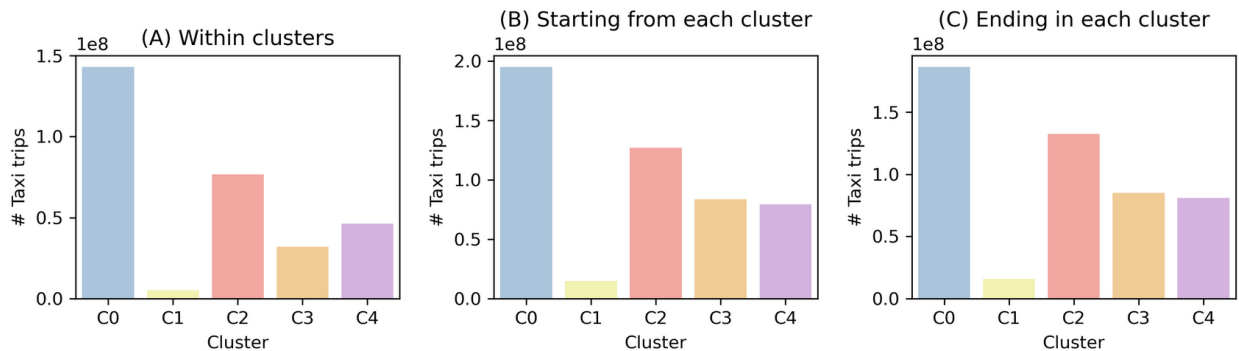

Figure 9. The number of taxi trips (A) within each cluster, (B) starting from each cluster regardless of destination clusters (C) ending in each of clusters regardless of origin clusters.

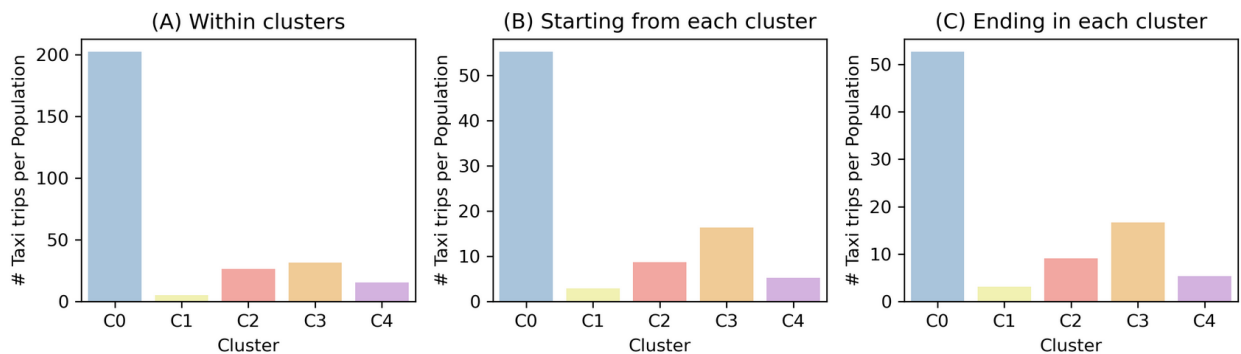

Figure 10. Normalized taxi trips per population (A) within each cluster, (B) starting from each cluster regardless of destination clusters (C) ending in each of clusters regardless of origin clusters.

Among all the neighborhood clusters, C1 (Suburban lifestyle in NYC) has the least amount of taxi trips originating from within. Located predominantly on the outskirts of NYC, C1 zones exhibit a suburban lifestyle. Notably, C1 zones have limited accessibility to public transit infrastructure, as there is only one subway line running in Staten Island, where the majority neighborhoods belong to C1 cluster. An important characteristic of C1 cluster is their high car ownership among households (see "noCarPercent" in Figure 5), a factor that significantly reduces dependence on both public transit (see comPublicTransit in Figure 5) and taxi or ride-sharing services (see "comTaxi" in Figure 5). In terms of commuting, C1 residents heavily rely on private vehicles, as driving alone and carpooling are the two most dominant commuting methods (see "comDriveAlone" and "comCarpool" in Figure 5). Figure 6 also highlights C1, reflecting NYC's suburban lifestyle, with the highest rate of solo driving and carpooling across neighborhoods. About 60% commute this way, 30% prefer public transit, and less than 5% choose biking or walking. In contrast, in neighborhoods like C0 (Elite Manhattan Resident), only 5% commute by driving or carpooling, 40% by public transit, and 27% by biking or walking. In addition, with a higher proportion of residents aged 60 and above (see "age>60Percent" in Figure 5), C1 residents tend to prioritize the convenience and familiarity offered by private vehicles. This demographic composition, together with suburban lifestyle of C1, explains the prevalence of privately owner vehicles and less reliance of taxi and ride-sharing services in C1 zones.

### 5.2.2. SafeGraph trip flows

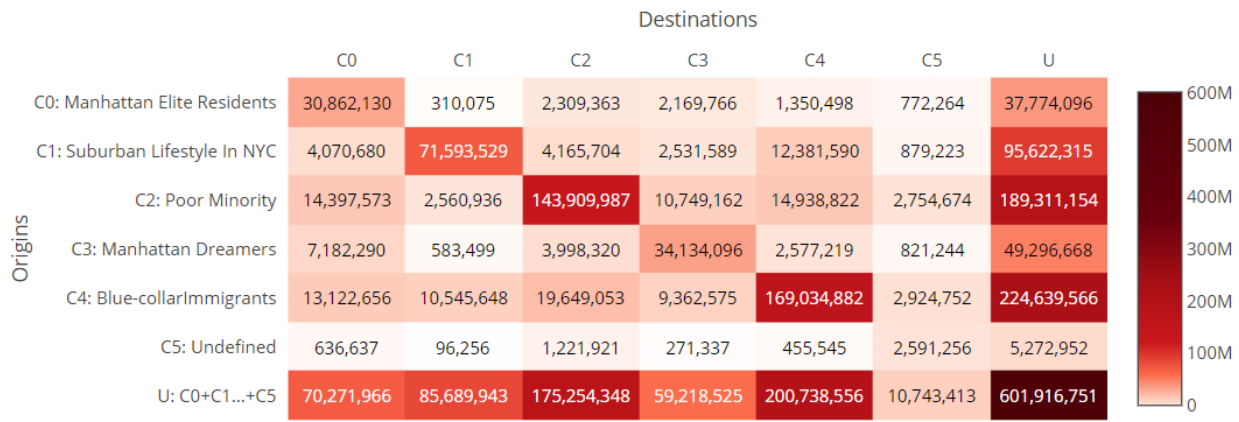

Figure 11. SafeGraph trip flows between clusters. U represents all clusters. For example, the cell in the bottom left corner represents the number of trips starting from any cluster and ending in C0. Similarly, the cell in the top right corner represents the number of trips starting from C0 and ending in any cluster.

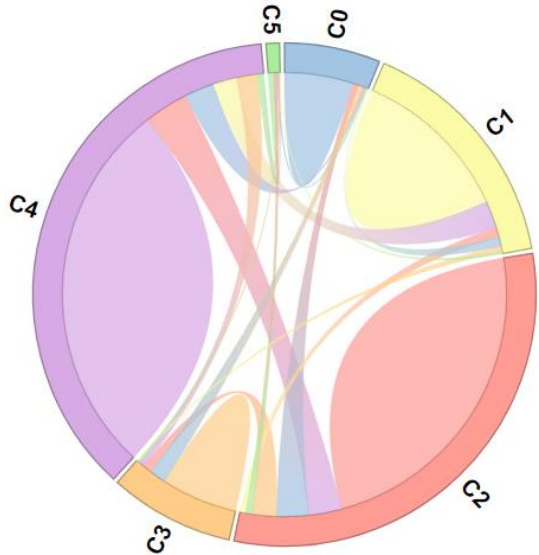

Figure 12. Chord diagram presenting the number of SafeGraph trips between clusters. C0: Elite Manhattan Residents; C1: Suburban Lifestyle in NYC; C2: Poor Minority; C3: Manhattan Dreamers; C4: Blue-collar Immigrants; C5: Undefined.

Figure 11 presents the origin-destination trip flows based on SafeGraph mobility patterns. Figure 12 presents a chord diagram for the OD trips. During the 27-month period, a total of 601,916,751 trips were recorded by SafeGraph. Among these clusters, C4 (Blue-collar Immigrants) has the largest number of trips. A total of 224,639,566 trips originated from a C4 (Blue-collar immigrants), which is about 37.3% of all the trips. In addition, 200,738,556 trips ended in a C4 (Blue-collar immigrants), consisting of 33.35% of all the trips. Among these, 169,034,882 trips occurred within C4, about 28.08% of all the SafeGraph trips. Next to C4 (Blue-collar immigrants), C2 (poor minority) zones have the second highest trip amount according to SafeGraph. It is origin for 189,311,154 trips and destination for 175,254,348 trips, consisting of 31.5% and 29.1%, respectively. Based on SafeGraph travel mobility patterns, among residential areas, C0 (Elite Manhattan residents) has the smallest amount of trips originated within, but C3 (Manhattan dreamers) has the smallest amount of trips ended within.

To comprehend the distinction between taxi and SafeGraph trips, it is essential to recognize that taxi trips encompass all trips taken via taxi services. Conversely, the number of SafeGraph trips is derived from mobile devices equipped with location-tracking applications. To understand the proportion of the population whose mobility is represented within SafeGraph data, we computed sample rates by using the method outlined by Li et al. (2024) in their prior study. Within the home panel summary data sourced from SafeGraph, there is a metric denoted as the "number_devices_residing" and "number_devices_primary_daytime". Each represents the number of distinct devices observed within a primary nighttime location and daytime location, respectively. This metric is the count of mobile devices equipped with location-tracking applications. Utilizing the available device count data at the block group level, we first calculated the average number of monthly devices during our study period from January 1st to March 31st,

2021. Then, we aggregated the average number of devices within each cluster as defined in Figure 4. Concurrently, we utilized the population data obtained from American Community Survey (ACS) and aggregated its counts at the block group level for each cluster. Then, we divided the number of distinct devices by the total population count. Table 2 shows the result.

Table 2. Sampling rate for SafeGraph in each neighborhood of NYC. (A) Total population count. The average number of monthly devices observed within primary nighttime locations (B) and daytime locations (C). (D) represents the sampling rate calculated using (B), while (E) represents the sampling rate calculated using (C). C0: Elite Manhattan Residents; C1: Suburban Lifestyle in NYC; C2: Poor Minority; C3: Manhattan Dreamers; C4: Blue-collar Immigrants; C5: Undefined.

| Cluster | (A) POP | (B) Devices Nighttime | (C) Devices Daytime | (D) Nighttime Sampling Rate | (E) Daytime Sampling Rate |
|---|---|---|---|---|---|
| C0 | 706357 | 81581.47 | 84205.76 | 11.55 % | 11.92 % |
| C1 | 1016627 | 65595.65 | 32439.22 | 6.45 % | 3.19 % |
| C2 | 2913214 | 168776.55 | 109351.70 | 5.79 % | 3.75 % |
| C3 | 1019873 | 45392.80 | 28786.16 | 4.45 % | 2.82 % |
| C4 | 3019832 | 167506.19 | 88181.50 | 5.55 % | 2.92 % |

Additionally, we examined the relationship among spatial distribution of devices, SafeGraph trips and population. Notably, we identified a robust correlation between the quantity of devices and the volume of trips; specifically, areas with a higher number of detected devices correspond to increased trip activity. Specifically, Pearson's correlation coefficient (r) stands at about 0.95. Furthermore, we found a correlation between the number of devices observed within nighttime and the number of populations identified from the American Community Survey. Specifically, Pearson's correlation coefficient (r) stands at approximately 0.8, indicating a positive relationship between these variables. In essence, the spatial distribution of trips closely mirrors that of the devices, and the spatial distribution of devices approximately mirrors that of the population.

Further insights were obtained by examining the ratio of SafeGraph Trips to the total number of devices. Using the device data, we compared the raw count of SafeGraph trips with the number of SafeGraph trips per device. Figure 13 illustrates the number of SafeGraph trips within each cluster (A), starting from each cluster (B), and ending in each cluster (C). In contrast, Figure 14 shows normalized SafeGraph trips per device within each cluster (A), starting from each cluster (B), and ending in each cluster (C). Across all three charts of Figure 13, C4 (Blue-collar Immigrants) emerges with the highest number of SafeGraph trips, while C1 (Suburban Lifestyle in NYC) records less than half of C4's count. However, in Figure 14, across all three cases (A), (B), and (C), C1 consistently showcases the highest number of SafeGraph trips per device. C1 is distinguished by a notable prevalence of car ownership, as illustrated in Figure 5, along with the lowest percentage of population utilizing public transit for commuting across neighborhoods, and the highest proportion of individuals commuting by driving alone, as highlighted in Figure 6. Consequently, a plausible explanation for the elevated number of SafeGraph trips per device depicted in C1 zone could stem from the likelihood that individuals who drive their own vehicles tend to utilize mobile applications with location tracking capabilities, such as Google Maps and Apple Maps. As they navigate using these apps installed on their mobile devices, SafeGraph collects their location data, resulting in an increased volume of trip records originating from, ending in, or traversing through C1 (Suburban Lifestyle in NYC)

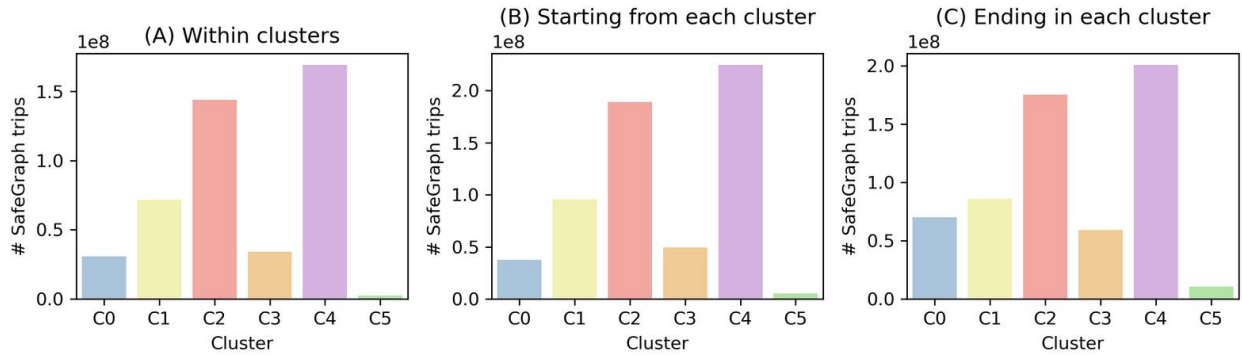

Figure 13. The number of SafeGraph trips (A) within each cluster, (B) starting from each cluster regardless of destination clusters (C) ending in each of clusters regardless of origin clusters.

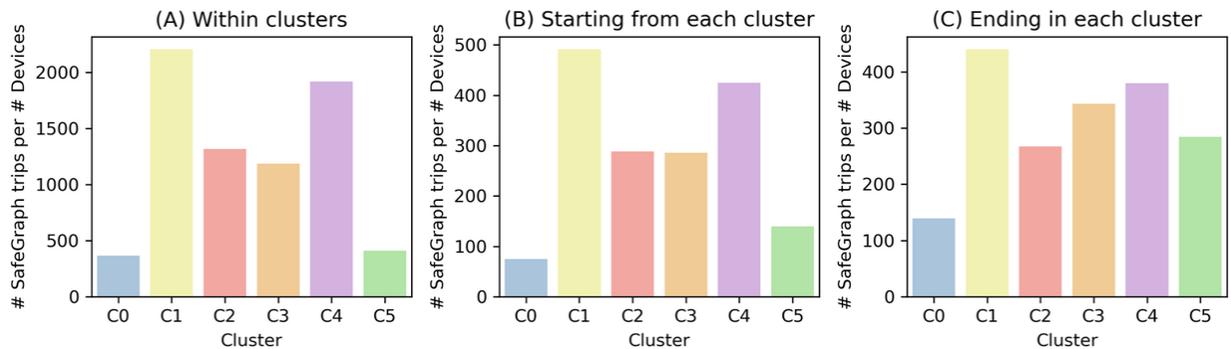

Figure 14. Normalized SafeGraph trips per device observed during daytime (A) within each cluster, (B) starting from each cluster regardless of destination clusters (C) ending in each of clusters regardless of origin clusters.

While C1 (Suburban Lifestyle in NYC) demonstrates a notable difference before and after normalization by the number of devices, as previously mentioned, C0 (Elite Manhattan Residents) consistently records either the lowest or the second lowest number of SafeGraph trips across all charts in Figures 13 and 14. One potential explanation for this observation could be attributed to the predominant nature of Points of Interest (POIs) where mobility data is collected. Typically, these locations encompass commercial establishments such as restaurants, retail stores, and grocery stores. However, C0 is characterized by the concentration of world-class companies, resulting in a higher prevalence of office spaces rather than consumer-oriented venues like restaurants and retail stores. Trips to and from company offices may be less likely to be recorded as trips in SafeGraph data. Moreover, C0 exhibits the smallest percentage of individuals commuting by car, as depicted in Figure 6. Residents in this area predominantly rely on walking, biking, or public transportation for their daily commute. As a result, individuals commuting via these modes are less inclined to use mobile applications with location tracking features, such as Google or Apple Maps, during their commute compared to those who commute by driving cars.

The absence of such phone activities may also contribute to the lower mobility data collected by SafeGraph in this area.

### 5.2.3. Comparison of Relative Frequency of SafeGraph and Taxi Trips

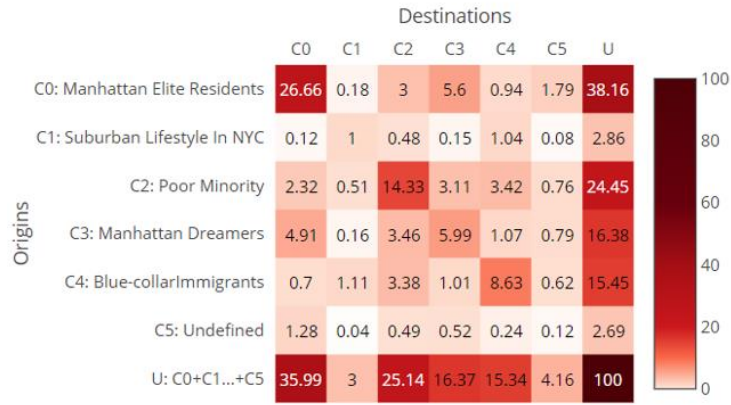

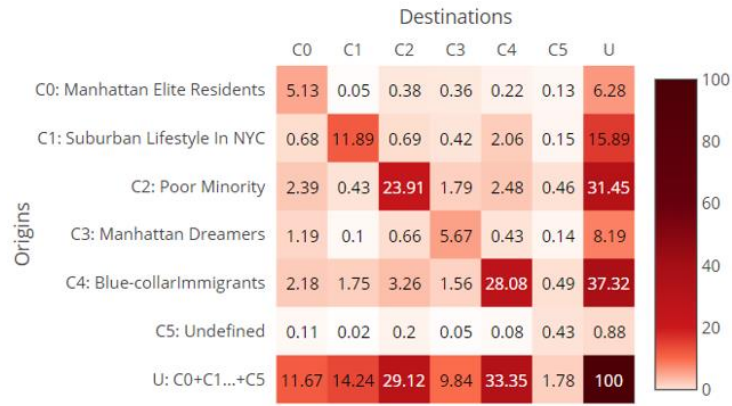

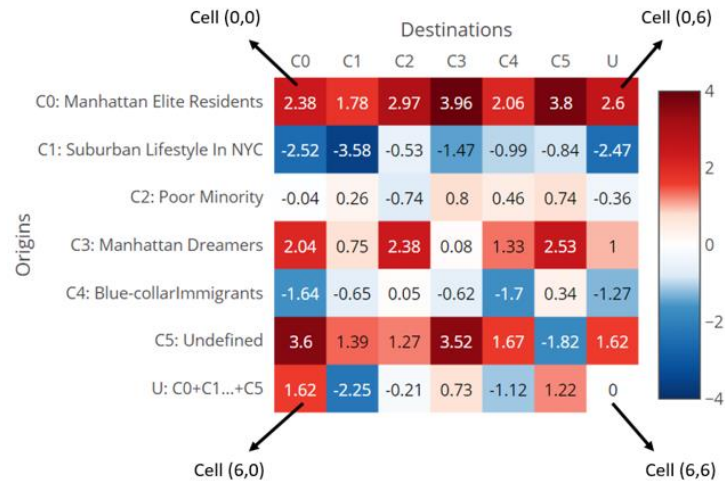

Figure 15. (A) The percentage of Taxi trips (B) The percentage of SafeGraph trips (C) Log of relative frequency ratio (LRFR). C0: Elite Manhattan Residents; C1: Suburban Lifestyle in NYC; C2: Poor Minority; C3: Manhattan Dreamers; C4: Blue-collar Immigrants; C5: Undefined. U represents all clusters. For example, the cell located in the bottom left corner in both (A) and (B) denotes the percentage of taxi and SafeGraph trips, respectively, originating from any cluster and ending in C0. The calculation of LRFR for the bottom left corner cell in (C) is based on the values of the corresponding cells in the bottom left corners of (A) and (B). The cell numbers for the four corner cells are indicated in the format Cell (x,y).

We analyzed the disparity between taxi trips and SafeGraph trips. Figure 15 (A) and (B) represent relative frequency for Taxi and SafeGraph, respectively. Relative frequency refers to the proportion or percentage of times a trip occurs relative to the total number of trips. (C) represents log of relative frequency ratio (LRFR) described in section 4.3. LRFR serves as a metric for comparing the relative frequency of Taxi trips (A) to that of SafeGraph Trips (B). In Figure 15 (C), red indicates a higher relative frequency of taxi trips compared to SafeGraph trips. This suggests that taxi usage as a mode of transportation is prominent relative to other means. Conversely, blue signifies a higher relative frequency of SafeGraph trips compared to taxi trips. This suggests that taxi usage is not predominant compared to other modes of transportation, such as driving, public transit, biking, or walking.

Trips associated with Manhattan, nearby regions, or LGA and JFK airports exhibit the top four LRFR of taxis to SafeGraph data. The highest taxi-to-SafeGraph LRFR was found for trips from C0 (Elite Manhattan Residents) to C3 (Manhattan Dreamers). The second largest LRFR was identified in trips from a C0 (Elite Manhattan Residents) zone to a C5 (Undefined zone), primarily to airports. Similarly, trips from a C5 (Undefined zone), mostly LGA and JFK airports to either a C0 (Elite Manhattan Residents) or a C3 (Manhattan Dreamers) also show 3rd and 4th largest taxi-to-SafeGraph LRFR.

The symmetrical pattern, with consistently high Taxi-to-SafeGraph LRFRs observed in both outgoing and incoming trips among (a) C0 (Elite Manhattan Residents) and C3 (Manhattan Dreamers), (b) C0 and C5 (mostly airports), and (c) C3 and C5, suggests that taxis serve as the primary mode of transportation for travel between these zones, regardless of the direction of the trip. Another aspect of symmetrical pattern is observed between C1 (Suburban Lifestyle in NYC) and C4 (Blue-collar Immigrants), indicating that the relative frequency of SafeGraph trips is larger than that taxi trips – taxis are not their dominant means of transportation between these two zones.

Unlike symmetrical patterns found above, asymmetrical patterns are also observed. For instance, flows from C1 (Suburban Lifestyle in NYC) to C0 (Elite Manhattan Residents) indicate significantly fewer relatively frequency of taxi trips compared to SafeGraph-recorded trips. Conversely, flows from C0 (Elite Manhattan Residents) to C1 (Suburban Lifestyle in NYC) show significantly greater relative frequency of taxi trips than SafeGraph trips. This inverse pattern indicates that residents mostly use transportation modes other than taxis, such as privately owned cars, for trips from C1 (Suburban Lifestyle in NYC) to C0 (Elite Manhattan Residents).

Conversely, when they travel from C0 to C1, they predominantly use taxis. This observation aligns with the low car ownership rate among residents in C0 (Elite Manhattan Residents) identified from Figure 5.

Furthermore, a similar inverse pattern is also evident in trips between C0 (Elite Manhattan Residents) and C4 (Blue-collar Immigrants). When traveling from C0 to C4, LRFR is 2.06, which means the relative frequency of taxi trips is approximately four (2 to the power of 2.06) times higher than that of SafeGraph trips. Conversely, when traveling from C4 to C0, LRFR is -1.64, the relative frequency of SafeGraph trips surpasses that of taxi trips by approximately threefold (2 to the power of 1.64). This finding suggests that individuals are more likely to opt for taxis when traveling from C0 to C4, whereas taxis are not the dominant mode of transportation for return trips from C4 to C0. This pattern can be understood by comparing transportation preferences between the two zones. C0 has the highest percentage of people commuting by taxi, whereas C4 has the second lowest (Figure 5). Furthermore, there is a notable difference in the percentage of people commuting by driving their cars, with approximately 3% in C0 and 38% in C4 (Figure 6). Based on these facts, we can infer that the abundance of taxis makes it easy for individuals to take taxis from C0 to C4. However, due to lower demand for taxis in C4, their availability is limited, making individuals less likely to use taxis when traveling back from C4 to C0.

Another noteworthy inverse pattern arises in trips between C1 (Suburban Lifestyle in NYC) and C5 (mostly airports). In this scenario, the LRFR of trips from C5 to C1 is 1.39, while the LRFR of trips from C1 to C5 is -0.84. These figures indicate a higher relative frequency of taxi usage when traveling from the airport to suburban areas compared to trips from suburban areas to the airport. This finding highlights a preference for taxis for inbound trips from the airport to suburban areas, likely due to the convenience of accessing taxis near airports compared to suburban areas, where taxi availability may be limited for journeys to the airport. The findings here are consistent with the results obtained from comparing SafeGraph trips originating from C5 (mostly airports) with those ending in C5 in Figure 14 (B) and (C) respectively. In contrast to the pattern observed in taxi trips, SafeGraph trips ending in C5 (airport regions) are nearly twice as high as those starting in C5. This disparity can be attributed to the high volume of individuals taking taxis at the airport, where taxis are readily accessible.

When examining the cluster-to-cluster trips, C0 (Elite Manhattan residents), as the origins, consistently presents high taxi-to-SafeGraph LRFR. In trips from C0 to all six destination neighborhoods, the LRFR of taxi trips consistently exceeds that of SafeGraph trips. Similarly, trips originating from C5 (Undefined zone) to the five neighborhood destinations, excluding itself, show high taxi-to-SafeGraph LRFR. Closer examination reveals that the top 100 LRFR for trips from C5 to all destination neighborhoods originate from C5 (Undefined zones), primarily starting from LGA airport.

It is worth noting that these two mobility datasets have major differences in terms of inclusiveness of trips. Taxi trip records encompass the majority of taxi and ride-sharing trips, except for those with GPS errors. SafeGraph, however, utilizes a sampling process, typically with less than a 10% sampling rate based on POI visitations. Notably, this rate is calculated based on the sampled number of devices over the total population, with the sampling rate for the number of visits remaining unknown. Additionally, both C0 (Elite Manhattan residents) and C3 (Manhattan dreamers) have convenient access to public transit, as these zones are primarily located in Manhattan's core or on the west side of Brooklyn and Queens, close to Manhattan. Consequently, there is a significant volume of trips originating from C0 or C3 zones conducted by public transit. Given these factors, a comparison using LRFR suggests that SafeGraph tends to undersample trips to/from C0 or C3.

## 6. Conclusion

To understand the representation of human mobility that can be captured by taxi and SafeGraph data, we compared taxi trip records with the number of trips captured by SafeGraph at the neighborhood level. To summarize the findings, taxi data typically records human mobility to and from areas with high taxi demand, such as the affluent districts from Lower to middle Manhattan, while often overlooking trip volumes from regions with lower taxi demand, notably in the suburbs of NYC. Conversely, SafeGraph data is effective at capturing trips to and from regions where many people rely on driving for their commute. These trips, often neglected by taxi data, are well represented in SafeGraph's dataset. Nonetheless, SafeGraph data tends to underestimate trips to and from Manhattan and its environs, where a considerable portion of the population depends on walking and public transportation for their daily commute. These results reflect the different transportation mode choices in different neighborhoods and phone usage among different groups of people in NYC.

This study presents the first direct comparative analysis of Taxi and SafeGraph data, offering novel insights into how these two widely used datasets capture human mobility patterns at the neighborhood scale. Although both datasets have been widely used independently to study human mobility, this study uniquely highlights their individual strengths, weaknesses, and inherent biases in capturing urban movement patterns. Through this comparative analysis, the study provides critical guidance on the applicability of each dataset for mobility research in different contexts.

A key contribution of this work is the introduction of a neighborhood-level analysis, which employs clustering based on demographic, socioeconomic, and commuting characteristics to reveal the relationship between these factors and mobility patterns identified in the Taxi and SafeGraph datasets. By categorizing neighborhoods and comparing the mobility patterns within each, the study uncovers patterns that would remain hidden.

Additionally, this research advances the field by introducing and applying the log of relative frequency ratio (LRFR) as a novel metric to detect transportation mode biases across regions or neighborhoods. The use of LRFR reveals that taxi data tends to overrepresent affluent areas, particularly in Manhattan, while SafeGraph data more effectively captures mobility in suburban, car-dependent regions. Notably, the LRFR metric is versatile and can be applied to compare any two human mobility datasets, such as those involving metro systems, air travel, or other transportation modes.

The study also offers significant insights into dataset biases and limitations. Taxi data shows a bias toward areas with high taxi demand, while SafeGraph tends to under-sample trips in pedestrian-heavy or public transit-dominated neighborhoods. This understanding of representativeness helps researchers and policymakers make more informed decisions about which dataset to use depending on the study's goals, particularly when capturing the movement of different population groups.

Importantly, the research emphasizes the necessity of cross-validation of multiple datasets in mobility studies. By highlighting the limitations of relying on a single data source, this study underscores the importance of using diverse datasets to obtain a more comprehensive and accurate picture of human mobility. Cross-referencing data from both Taxi and SafeGraph ensures that underrepresented or overrepresented population groups are properly accounted for in mobility analyses.

Although the comparison results generate insightful findings in possible sampling biases with SafeGraph mobility pattern dataset, it is crucial to acknowledge the limitations inherent in this study. First, our analysis focused on the number of trips, overlooking other important information associated with the trips, such as the traveling distance, the cost of different transportation modes, and the time of traveling. This limitation might have led to an oversimplification of the mobility patterns. For example, zones within the same cluster can be right next to each other or on the other side of the city. Future research should explore different travel measurements to achieve a more comprehensive understanding of the mobility patterns.

 Secondly, the demographic, socioeconomic, and commuting variables used in the analysis are derived from ACS, which is based on respondents' residential locations (where people live). However, human movement patterns across the city are more complex than just residential locations. Other factors—such as where people work, attend school, shop, or engage in recreational activities—also influence daily travel behavior. To gain a more complete understanding of urban mobility, future research should incorporate data that accounts for these additional factors. For example, point-of-interest (POI) data, which provides information on locations like businesses, schools, and entertainment venues, could be used to better understand why and where people move during the day, beyond their residential areas.

Thirdly, the number of trips were aggregated in this study, both spatially and temporally. The outcomes derived from this aggregated analysis might obscure understandings at a finer-grained level. This is especially true since the time span includes the Covid-19 pandemic. As travel behaviors underwent notable changes during various stages of the pandemic, summarizing the 27-month trip amount cannot capture such temporal variations. Recognizing this limitation, future study can investigate such patterns at a more detailed temporal level. However, the spatial resolution was limited by the original data provider, especially the taxi dataset. The spatial variations at a more detailed level may be harder to retain.

**Reference:**
Anda, C., Erath, A., & Fourie, P. J. (2017). Transport modelling in the age of big data. *International Journal of Urban Sciences*, *21*(sup1), 19–42.
Benita, F. (2021). Human mobility behavior in COVID-19: A systematic literature review and bibliometric analysis. *Sustainable Cities and Society*, *70*, 102916.
Bi, H., Ye, Z., Wang, C., Chen, E., Li, Y., & Shao, X. (2020). How built environment impacts online car-hailing ridership. *Transportation Research Record*, *2674*(8), 745–760.
Boarnet, M., & Crane, R. (2001). The influence of land use on travel behavior: specification and estimation strategies. *Transportation Research Part A: Policy and Practice*, *35*(9), 823-845.
Böcker, L., Dijst, M., & Faber, J. (2016). Weather, transport mode choices and emotional travel experiences. *Transportation Research Part A: Policy and Practice*, *94*, 360–373.
Buehler, R. (2011). Determinants of transport mode choice: A comparison of Germany and the USA. *Journal of Transport Geography*, *19*(4), 644–657.
Ceder, A. (2021). Urban mobility and public transport: Future perspectives and review. *International Journal of Urban Sciences*, *25*(4), 455–479.
Chang, S., Pierson, E., Koh, P. W., Gerardin, J., Redbird, B., Grusky, D., & Leskovec, J. (2021). Mobility network models of COVID-19 explain inequities and inform reopening. *Nature*, *589*(7840), 82–87.
Chen, H., Tao, F., Ma, P., Gao, L., & Zhou, T. (2021). Applicability evaluation of several spatial clustering methods in spatiotemporal data mining of floating car trajectory. *ISPRS International Journal of Geo-Information*, *10*(3), 161.
Chen, P., Zhai, W., & Yang, X. (2023). Enhancing resilience and mobility services for vulnerable groups facing extreme weather: Lessons learned from Snowstorm Uri in Harris County, Texas. *Natural Hazards*, *118*(2), 1573–1594.
Choi, J., No, W., Park, M., & Kim, Y. (2022). Inferring land use from spatialtemporal taxi ride data. *Applied Geography*, *142*, 102688.
Chuah, S. P., Wu, H., Lu, Y., Yu, L., & Bressan, S. (2016). *Bus routes design and optimization via taxi data analytics*. 2417–2420.
Colombo, C., Borri, C. (2020). Human mobility insights for urban planners: A SafeGraph case study. *Transportation Research Procedia*, 45, 528-537.
Convery, S., & Williams, B. (2019). Determinants of Transport Mode Choice for Non-Commuting Trips: The Roles of Transport, Land Use and Socio-Demographic Characteristics. *Urban Science*, *3*(3), 82.
Coston, A., Guha, N., Ouyang, D., Lu, L., Chouldechova, A., & Ho, D. E. (2021). *Leveraging administrative data for bias audits: Assessing disparate coverage with mobility data for COVID-19 policy*. 173–184.
Dargin, J. S., Li, Q., Jawer, G., Xiao, X., & Mostafavi, A. (2021). Compound hazards: An examination of how hurricane protective actions could increase transmission risk of COVID-19. *International Journal of Disaster Risk Reduction*, *65*, 102560.
Delmelle, E. C. (2016). Mapping the DNA of urban neighborhoods: Clustering longitudinal sequences of neighborhood socioeconomic change. *Annals of the American Association of Geographers*, *106*(1), 36–56.
Delmelle, E. C. (2017). Differentiating pathways of neighborhood change in 50 US metropolitan areas. *Environment and Planning A*, *49*(10), 2402–2424.

Dokuz, A. S. (2022). Weighted spatio-temporal taxi trajectory big data mining for regional traffic estimation. *Physica A: Statistical Mechanics and Its Applications*, *589*, 126645.
Foote, N., & Walter, R. (2017). Neighborhood and socioeconomic change in emerging megapolitan nodes: Tracking shifting social geographies in three rapidly growing United States metropolitan areas, 1980–2010. *Urban Geography*, *38*(8), 1203–1230.
Gebeyehu, M., & Takano, S. (2007). Diagnostic evaluation of public transportation mode choice in Addis Ababa. *Journal of Public Transportation*, *10*(4), 27–50.
Ghaffar, A., Mitra, S., & Hyland, M. (2020). Modeling determinants of ridesourcing usage: A census tract-level analysis of Chicago. *Transportation Research Part C: Emerging Technologies*, *119*, 102769.
Gupta, S., Simon, K., & Wing, C. (2020). The impact of state and local policies on human mobility patterns in the United States during the COVID-19 pandemic. *Journal of Public Economics*, 191, 104254.
Han, S. Y., Kang, J., Lyu, F., Baig, F., Park, J., Smilovsky, D., & Wang, S. (2023). A cyberGIS approach to exploring neighborhood-level social vulnerability for disaster risk management. *Transactions in GIS*, *27*(7), 1942–1958.
Handley, M. Z., & Blumenthal, E. (2021). Tracking the effects of COVID-19 interventions on human mobility using SafeGraph data. *Scientific Reports*, 11(1), 9372.
He, T., Bao, J., Li, R., Ruan, S., Li, Y., Song, L., He, H., & Zheng, Y. (2020). *What is the human mobility in a new city: Transfer mobility knowledge across cities*. 1355–1365.
Hsu, C.-W., Liu, C., Nguyen, K. M., Chien, Y.-H., & Mostafavi, A. (2024). Do human mobility network analyses produced from different location-based data sources yield similar results across scales? *Computers, Environment and Urban Systems*, *107*, 102052.
Hu, T., Wang, S., She, B., Zhang, M., Huang, X., Cui, Y., Khuri, J., Hu, Y., Fu, X., & Wang, X. (2021). Human mobility data in the COVID-19 pandemic: Characteristics, applications, and challenges. *International Journal of Digital Earth*, *14*(9), 1126–1147.
Huang, X., Jiang, Y., & Mostafavi, A. (2024). The emergence of urban heat traps and human mobility in 20 US cities. *Npj Urban Sustainability*, *4*(1), 6.
Huang, X., Li, Z., Jiang, Y., Li, X., & Porter, D. (2020). Twitter reveals human mobility dynamics during the COVID-19 pandemic. *PloS One*, *15*(11), e0241957.
Huang, X., Li, Z., Lu, J., Wang, S., Wei, H., & Chen, B. (2020). Time-series clustering for home dwell time during COVID-19: What can we learn from it? *ISPRS International Journal of Geo-Information*, *9*(11), 675.
Isaacman, S., Becker, R., Cáceres, R., Martonosi, M., Rowland, J., Varshavsky, A., & Willinger, W. (2012). *Human mobility modeling at metropolitan scales*. 239–252.
Liang, Y., Zhang, S., Zou, L., & Zheng, Y. (2018). Predicting citywide traffic using deep learning. *In Proceedings of the 26th ACM SIGSPATIAL International Conference on Advances in Geographic Information Systems*, 157-165.
Liu, Y., Liu, X., Gao, S., Gong, L., Kang, C., Zhi, Y., ... & Shi, L. (2015). Social sensing: A new approach to understanding our socioeconomic environments. *Annals of the Association of American Geographers*, *105*(3), 512-530.
Jay, J., Heykoop, F., Hwang, L., Courtepatte, A., de Jong, J., & Kondo, M. (2022). Use of smartphone mobility data to analyze city park visits during the COVID-19 pandemic. *Landscape and Urban Planning*, *228*, 104554.
Jiang, Y., Guo, D., Li, Z., & Hodgson, M. E. (2021). A novel big data approach to measure and visualize urban accessibility. *Computational Urban Science*, *1*(1), 1–15.

Jiang, Y., Huang, X., & Li, Z. (2021). Spatiotemporal Patterns of Human Mobility and Its Association with Land Use Types during COVID-19 in New York City. *ISPRS International Journal of Geo-Information*, *10*(5), 344.

Jiang, Y., Li, Z., & Ye, X. (2019). Understanding demographic and socioeconomic biases of geotagged Twitter users at the county level. *Cartography and Geographic Information Science*, *46*(3), 228–242.

Jiang, Y., Popov, A. A., Li, Z., & Hodgson, M. E. (2022). An optimal sensors-based simulation method for spatiotemporal event detection. *arXiv Preprint arXiv:2208.07969*.

Jiang, Y., Yuan, F., Farahmand, H., Acharya, K., Zhang, J., & Mostafavi, A. (2023). Data-driven tracking of the bounce-back path after disasters: Critical milestones of population activity recovery and their spatial inequality. *International Journal of Disaster Risk Reduction*, *92*, 103693.

Juhász, L., & Hochmair, H. H. (2020). *Studying spatial and temporal visitation patterns of points of interest using SafeGraph data in Florida*.

Kang, W., Rey, S., Wolf, L., Knaap, E., & Han, S. (2020). Sensitivity of sequence methods in the study of neighborhood change in the United States. *Computers, Environment and Urban Systems*, *81*, 101480.

Kang, Y., Gao, S., Liang, Y., Li, M., Rao, J., & Kruse, J. (2020). Multiscale dynamic human mobility flow dataset in the US during the COVID-19 epidemic. *Scientific Data*, *7*(1), 1–13.

Kodinariya, T. M., & Makwana, P. R. (2013). Review on determining number of Cluster in K-Means Clustering. *International Journal*, *1*(6), 90–95.

Kong, X., Xu, Z., Shen, G., Wang, J., Yang, Q., & Zhang, B. (2016). Urban traffic congestion estimation and prediction based on floating car trajectory data. *Future Generation Computer Systems*, *61*, 97–107.

Kuang, W., An, S., & Jiang, H. (2015). Detecting traffic anomalies in urban areas using taxi GPS data. *Mathematical Problems in Engineering*, *2015*.

Lessani, M. N., Li, Z., Jing, F., Qiao, S., Zhang, J., Olatosi, B., & Li, X. (2023). Human mobility and the infectious disease transmission: A systematic review. *Geo-Spatial Information Science*, 1–28.

Li, B., Cai, Z., Jiang, L., Su, S., & Huang, X. (2019). Exploring urban taxi ridership and local associated factors using GPS data and geographically weighted regression. *Cities*, *87*, 68–86.

Li, H., Xu, X., Li, X., Ma, S., & Zhang, H. (2021). Characterizing the urban spatial structure using taxi trip big data and implications for urban planning. *Frontiers of Earth Science*, *15*, 70–80.

Li, T., Jing, P., Li, L., Sun, D., & Yan, W. (2019). Revealing the varying impact of urban built environment on online car-hailing travel in spatio-temporal dimension: An exploratory analysis in Chengdu, China. *Sustainability*, *11*(5), 1336.

Li, Z., Li, X., Porter, D., Zhang, J., Jiang, Y., Olatosi, B., & Weissman, S. (2020). Monitoring the Spatial Spread of COVID-19 and Effectiveness of Control Measures Through Human Movement Data: Proposal for a Predictive Model Using Big Data Analytics. *JMIR Research Protocols*, *9*(12), e24432.

Li, Z., Ning, H., Jing, F., & Lessani, M. N. (2024). Understanding the bias of mobile location data across spatial scales and over time: A comprehensive analysis of SafeGraph data in the United States. *Plos One*, *19*(1), e0294430.

Liang, Y., Yin, J., Pan, B., Lin, M. S., Miller, L., Taff, B. D., & Chi, G. (2022). Assessing the validity of mobile device data for estimating visitor demographics and visitation patterns in Yellowstone National Park. *Journal of Environmental Management*, *317*, 115410. Lin, Y., Li, W., Qiu, F., & Xu, H. (2012). Research on optimization of vehicle routing problem for ride-sharing taxi. *Procedia-Social and Behavioral Sciences*, *43*, 494–502.
Ling, C., & Delmelle, E. C. (2016). Classifying multidimensional trajectories of neighbourhood change: A self-organizing map and k-means approach. *Annals of GIS*, *22*(3), 173–186.
Liu, C., Wang, S., Cuomo, S., & Mei, G. (2021). Data analysis and mining of traffic features based on taxi GPS trajectories: A case study in Beijing. *Concurrency and Computation: Practice and Experience*, *33*(3), e5332.
Liu, Y., Kang, C., Gao, S., Xiao, Y., & Tian, Y. (2012). Understanding intra-urban trip patterns from taxi trajectory data. *Journal of Geographical Systems*, *14*(4), 463–483.
Lu, Y., & Giuliano, G. (2023). Understanding mobility change in response to COVID-19: A Los Angeles case study. *Travel Behaviour and Society*, *31*, 189–201.
Luca, M., Barlacchi, G., Lepri, B., & Pappalardo, L. (2021). A survey on deep learning for human mobility. *ACM Computing Surveys (CSUR)*, *55*(1), 1–44.
Lyu, T., Wang, P. S., Gao, Y., & Wang, Y. (2021). Research on the big data of traditional taxi and online car-hailing: A systematic review. *Journal of Traffic and Transportation Engineering (English Edition)*, *8*(1), 1–34.
Ma, X., Wang, Z., Chen, F., & Huang, J. (2015). Deep learning for traffic flow prediction. *IEEE Transactions on Intelligent Transportation Systems*, 16(2), 865-873.
Malik, M. M., Lamba, H., Nakos, C., & Pfeffer, J. (2015). Population bias in geotagged tweets. *People*, *1*(3,759.710), 3–759.
Markou, I., Rodrigues, F., & Pereira, F. C. (2017). Use of taxi-trip data in analysis of demand patterns for detection and explanation of anomalies. *Transportation Research Record*, *2643*(1), 129-138.
Manson, S., Schroeder, J., Van Riper, D., Knowles, K., Kugler, T., Roberts, F., & Ruggles, S. (2023). *IPUMS National Historical Geographic Information System: Version 18.0 [dataset]* [dataset]. Minneapolis, MN: IPUMS. https://doi.org/10.18128/D050.V18.0
Mislove, A., Lehmann, S., Ahn, Y.-Y., Onnela, J.-P., & Rosenquist, J. (2011). *Understanding the demographics of Twitter users*. *5*(1), 554–557.
Mukherjee, S., & Jain, T. (2022). Impact of COVID-19 on the mobility patterns: An investigation of taxi trips in Chicago. *Plos One*, *17*(5), e0267436.
Ning, H., Li, Z., Qiao, S., Zeng, C., Zhang, J., Olatosi, B., & Li, X. (2023). Revealing geographic transmission pattern of COVID-19 using neighborhood-level simulation with human mobility data and SEIR model: A Case Study of South Carolina. *International Journal of Applied Earth Observation and Geoinformation*, *118*, 103246.
NYC TLC. (n.d.). *TLC Trip Record Data—TLC*. Retrieved February 22, 2024, from https://www.nyc.gov/site/tlc/about/tlc-trip-record-data.page
Pepe, E., Bajardi, P., Gauvin, L., Privitera, F., Lake, B., Cattuto, C., & Tizzoni, M. (2020). COVID-19 outbreak response, a dataset to assess mobility changes in Italy following national lockdown. *Scientific Data*, *7*(1), 230.
Pike, S., & Lubell, M. (2016). Geography and social networks in transportation mode choice. *Journal of Transport Geography*, *57*, 184–193.
Qian, X., & Ukkusuri, S. V. (2015). Spatial variation of the urban taxi ridership using GPS data. *Applied Geography*, *59*, 31–42.

Qu, B., Yang, W., Cui, G., & Wang, X. (2019). Profitable taxi travel route recommendation based on big taxi trajectory data. *IEEE Transactions on Intelligent Transportation Systems*, *21*(2), 653–668.
Pan, G., Qi, G., Wu, Z., Zhang, D., & Li, S. (2012). Land-use classification using taxi GPS traces. *IEEE Transactions on Intelligent Transportation Systems*, *14*(1), 113-123.
Parker, J. C., Neumann, R., & Brown, K. F. (2021). Foot traffic and the fiscal impacts of COVID-19: Evidence from SafeGraph data. *Regional Science and Urban Economics*, 91, 103708.
Peng, C., Jin, X., Wong, K. C., Shi, M., & Liò, P. (2012). Collective human mobility pattern from taxi trips in urban area. *PloS one*, *7*(4), e34487.
Rahman, M. M., Mou, J. R., Tara, K., & Sarkar, M. I. (2016). *Real time Google map and Arduino based vehicle tracking system*. 1–4.
Ranjan, G., Zang, H., Zhang, Z.-L., & Bolot, J. (2012). Are call detail records biased for sampling human mobility? *ACM SIGMOBILE Mobile Computing and Communications Review*, *16*(3), 33–44.
Rey, S. J., Anselin, L., Folch, D. C., Arribas-Bel, D., Sastré Gutiérrez, M. L., & Interlante, L. (2011). Measuring spatial dynamics in metropolitan areas. *Economic Development Quarterly*, *25*(1), 54–64.
Rey, S., Knaap, E., Han, S., Wolf, L., & Kang, W. (2018). *Spatio-temporal analysis of socioeconomic neighborhoods: The Open Source Longitudinal Neighborhood Analysis Package (OSLNAP)*. Proceedings of the Python in Science Conference.
Riascos, A., & Mateos, J. L. (2020). Networks and long-range mobility in cities: A study of more than one billion taxi trips in New York City. *Scientific Reports*, *10*(1), 4022.
Ricord, S., & Wang, Y. (2023). Investigation of equity biases in transportation data: A literature review synthesis. *Journal of Transportation Engineering, Part A: Systems*, *149*(11), 03123004.
SafeGraph. (2020). *Social Distancing Metrics*. SafeGraph. https://docs.safegraph.com/docs/social-distancing-metrics
Santi, P., Resta, G., Szell, M., Sobolevsky, S., Strogatz, S. H., & Ratti, C. (2014). Quantifying the benefits of vehicle pooling with shareability networks. *Proceedings of the National Academy of Sciences*, 111(37), 13290-13294.
Scholz, R. W., & Lu, Y. (2014). Detection of dynamic activity patterns at a collective level from large-volume trajectory data. International Journal of Geographical Information Science, 28(5), 946-963.
Shaw, S.-L., Tsou, M.-H., & Ye, X. (2016). Editorial: Human dynamics in the mobile and big data era. *International Journal of Geographical Information Science*, *30*(9), 1687–1693. https://doi.org/10.1080/13658816.2016.1164317
Shen, J., Liu, X., & Chen, M. (2017). Discovering spatial and temporal patterns from taxi-based Floating Car Data: A case study from Nanjing. *GIScience & Remote Sensing*, *54*(5), 617–638.
Siangsuebchart, S., Ninsawat, S., Witayangkurn, A., & Pravinvongvuth, S. (2021). Public transport gps probe and rail gate data for assessing the pattern of human mobility in the bangkok metropolitan region, Thailand. *Sustainability*, *13*(4), 2178.
Siła-Nowicka, K., Vandrol, J., Oshan, T., Long, J. A., Demšar, U., & Fotheringham, A. S. (2016). Analysis of human mobility patterns from GPS trajectories and contextual information. *International Journal of Geographical Information Science*, *30*(5), 881–906.

Squire, R. F. (2019, October 17). *What about bias in the SafeGraph dataset?* https://www.safegraph.com/blog/what-about-bias-in-the-safegraph-dataset

Susswein, Z., Rest, E. C., & Bansal, S. (2023). Disentangling the rhythms of human activity in the built environment for airborne transmission risk: An analysis of large-scale mobility data. *Elife*, *12*.

Szymanski, P., & Malinowski, K. (2020). A scalable dynamic network model for mobility tracking. *IEEE Access*, 8, 153749-153762.

Takahashi, N., Miyamoto, S., & Asano, M. (2004). *Using taxi GPS to gather high-quality traffic data for winter road management evaluation in Sapporo, Japan*. 455–469.

Tang, J., Liu, F., Wang, Y., & Wang, H. (2015). Uncovering urban human mobility from large scale taxi GPS data. *Physica A: Statistical Mechanics and Its Applications*, *438*, 140–153.

Tang, J., Zhu, Y., Huang, Y., Peng, Z.-R., & Wang, Z. (2019). Identification and interpretation of spatial–temporal mismatch between taxi demand and supply using global positioning system data. *Journal of Intelligent Transportation Systems*, *23*(4), 403–415.

*TLC Trip Records User Guide*. (2019, September 23).

Tu, W., Cao, R., Yue, Y., Zhou, B., Li, Q., & Li, Q. (2018). Spatial variations in urban public ridership derived from GPS trajectories and smart card data. *Journal of Transport Geography*, *69*, 45–57.

Vachuska, K. (2024). Cold weather isolation is worse in poor and non-white neighborhoods in the United States. *Preventive Medicine Reports*, *38*, 102541.

Wang, A., Zhang, A., Chan, E. H., Shi, W., Zhou, X., & Liu, Z. (2020). A review of human mobility research based on big data and its implication for smart city development. *ISPRS International Journal of Geo-Information*, *10*(1), 13.

Wang, J., McDonald, N., Cochran, A. L., Oluyede, L., Wolfe, M., & Prunkl, L. (2021). Health care visits during the COVID-19 pandemic: A spatial and temporal analysis of mobile device data. *Health & Place*, *72*, 102679.

Wang, Z., Zhang, Y., Jia, B., & Gao, Z. (2024). Comparative Analysis of Usage Patterns and Underlying Determinants for Ride-hailing and Traditional Taxi Services: A Chicago Case Study. *Transportation Research Part A: Policy and Practice*, *179*, 103912.

Wei, F., & Knox, P. L. (2014). Neighborhood change in metropolitan America, 1990 to 2010. *Urban Affairs Review*, *50*(4), 459–489.

Wei, R., Rey, S., & Knaap, E. (2021). Efficient regionalization for spatially explicit neighborhood delineation. *International Journal of Geographical Information Science*, *35*(1), 135–151.

Wei, Z., & Mukherjee, S. (2023). Examining income segregation within activity spaces under natural disaster using dynamic mobility network. *Sustainable Cities and Society*, *91*, 104408.

Weill, J. A., Stigler, M., Deschenes, O., & Springborn, M. R. (2020). Social distancing responses to COVID-19 emergency declarations strongly differentiated by income. *Proceedings of the National Academy of Sciences*, *117*(33), 19658–19660.

Wesolowski, A., Eagle, N., Noor, A. M., Snow, R. W., & Buckee, C. O. (2013). The impact of biases in mobile phone ownership on estimates of human mobility. *Journal of the Royal Society Interface*, *10*(81), 20120986.

Willis, G., & Tranos, E. (2021). Using 'Big Data' to understand the impacts of Uber on taxis in New York City. *Travel Behaviour and Society*, *22*, 94–107.

Yang, J., Yu, M., Qin, H., Lu, M., & Yang, C. (2019). A twitter data credibility framework—Hurricane Harvey as a use case. *ISPRS International Journal of Geo-Information*, *8*(3), 111.

Yang, Z., Franz, M. L., Zhu, S., Mahmoudi, J., Nasri, A., & Zhang, L. (2018). Analysis of Washington, DC taxi demand using GPS and land-use data. *Journal of Transport Geography*, *66*, 35–44.

Yao, X., Zhu, D., Gao, Y., Wu, L., Zhang, P., & Liu, Y. (2018). A stepwise spatio-temporal flow clustering method for discovering mobility trends. Ieee Access, 6, 44666-44675.

Yu, W., Wang, W., Hua, X., & Wei, X. (2021). Exploring taxi demand distribution of comprehensive land-use patterns using online car-hailing data and points of interest in Chengdu, China. *Transportation Research Record*, *2675*(10), 1268-1286.

Yuan, Y., Lu, Y., Chow, T. E., Ye, C., Alyaqout, A., & Liu, Y. (2021). The missing parts from social media-enabled smart cities: Who, where, when, and what? In *Smart Spaces and Places* (pp. 130–142). Routledge.

Zhang, B., Chen, S., Ma, Y., Li, T., & Tang, K. (2020). Analysis on spatiotemporal urban mobility based on online car-hailing data. *Journal of Transport Geography*, *82*, 102568.

Zhang, D., Huang, J., Li, Y., Zhang, F., Xu, C., & He, T. (2014). *Exploring human mobility with multi-source data at extremely large metropolitan scales*. 201–212.

Zhang, L., Yang, F., Schwanen, T., & Handy, S. (2020). The impact of the COVID-19 pandemic on human mobility and daily travel: A country-wide analysis using SafeGraph data. *Transportation Research Part A: Policy and Practice*, 144, 1-17

Zhang, W., Qi, G., Pan, G., Lu, H., Li, S., & Wu, Z. (2015). City-scale social event detection and evaluation with taxi traces. *ACM Transactions on Intelligent Systems and Technology (TIST)*, *6*(3), 1-20.

Zhang, X., Huang, B., & Zhu, S. (2019). Spatiotemporal influence of urban environment on taxi ridership using geographically and temporally weighted regression. *ISPRS International Journal of Geo-Information*, *8*(1), 23.

Zhao, K., Tarkoma, S., Liu, S., & Vo, H. (2016). *Urban human mobility data mining: An overview*. 1911–1920.

Zhao, Z., Shaw, S.-L., Xu, Y., Lu, F., Chen, J., & Yin, L. (2016). Understanding the bias of call detail records in human mobility research. *International Journal of Geographical Information Science*, *30*(9), 1738–1762.

Zheng, Y., Li, Q., Chen, Y., Xie, X., & Ma, W.-Y. (2008). Understanding mobility based on GPS data. *Proceedings of the 10th International Conference on Ubiquitous Computing*, 312–321.

Zhu, X., & Guo, D. (2014). Mapping large spatial flow data with hierarchical clustering. Transactions in GIS, 18(3), 421-435.

Zhu, X., & Guo, D. (2017). Urban event detection with big data of taxi OD trips: A time series decomposition approach. *Transactions in GIS*, *21*(3), 560–574.